\documentclass[aps,prd,amssymb,amsmath,nobibnotes,nofootinbib,superscriptaddress, reprint, 12pt ]{revtex4}
\usepackage{amsfonts,amsmath,hyperref,url, color}
\allowdisplaybreaks
\usepackage{bm, bbm}
\usepackage{graphicx}
\usepackage{mathtools}
\usepackage{t1enc}
\usepackage{hepnames}
\usepackage{tensor}
\usepackage{comment}

\usepackage{dcolumn}

\usepackage{adjustbox,booktabs}

\usepackage{makecell}
\usepackage{array}

\def\bbbone{{\mathchoice {\rm 1\mskip-4mu l} {\rm 1\mskip-4mu l}
{\rm 1\mskip-4.5mu l} {\rm 1\mskip-5mu l}}}

\usepackage{soul}

\newcommand{\bc}{\begin{center}}
\newcommand{\ec}{\end{center}}
\newcommand{\be}{\begin{eqnarray}}
\newcommand{\ee}{\end{eqnarray}}
\newcommand{\bs}{\begin{slide}}
\newcommand{\es}{\end{slide}}

\newcommand{\bi}{\begin{itemize}}
\newcommand{\ei}{\end{itemize}}

\def\bbbone{{\mathchoice {\rm 1\mskip-4mu l} {\rm 1\mskip-4mu l}
{\rm 1\mskip-4.5mu l} {\rm 1\mskip-5mu l}}}

\begin{document}
\title{$\kappa$-deformed complex scalar field: conserved charges, symmetries and their impact on physical observables}

\author{Andrea Bevilacqua}
\email{andrea.bevilacqua@ncbj.gov.pl}
\affiliation{National Centre for Nuclear Research, ul. Pasteura 7, 02-093 Warsaw, Poland}
\author{Jerzy Kowalski-Glikman}
\email{jerzy.kowalski-glikman@uwr.edu.pl}
\affiliation{University of Wroc\l{}aw, Faculty of Physics and Astronomy, pl.\ M.\ Borna 9, 50-204
Wroc\l{}aw, Poland}
\affiliation{National Centre for Nuclear Research, ul. Pasteura 7, 02-093 Warsaw, Poland}
\author{Wojciech Wi\'slicki}
\email{wojciech.wislicki@ncbj.gov.pl}
\affiliation{National Centre for Nuclear Research, ul. Pasteura 7, 02-093 Warsaw, Poland}

\date{\today}

\begin{abstract}
In this paper we revisit the model of $\kappa$-deformed complex scalar field introduced in our previous paper \cite{Arzano:2020jro}. We find that the model possesses ten conserved Noether charges that form, under commutators, a representation of (undeformed) Poincar\'e algebra, proving that it is relativistic and does not break Lorentz invariance. It turns out, however that the spacetime representation of boosts is not standard and contains a non-local translation, different for positive and negative energy modes. It then follows that although the masses of particles and anti-particles are equal, the model violates  CPT symmetry in a subtle way. We explain why the Jost-Wightman-Greenberg theorem of equivalence of the Poincar\'e symmetry and  CPT fails in our case. Finally, we discuss the phenomenological consequences of the theory and its possible observational signatures.

\end{abstract}`

\maketitle

\section{Introduction}

Spacetime symmetries are the single most important organizing principle of modern fundamental physics. In the context of Minkowski space field theories we have to do with Poincar\'e symmetry that  reflects the  basic properties of flat space-time  codified in special relativity, i.e. the fact that in such space-time no direction and no point are special. This fact is related to the properties of inertial observers and relativity principle: any translated, boosted or rotated inertial observer should describe the physics of a process in an equivalent way. Another important consequence of the Poincar\'e symmetry is that, as a consequence of the Noether theorem, we have to do with a number of conserved quantities (energy, momentum, angular momentum), which for isolated systems do not change in time. In the  canonical quantum field theory the charges are expressed as quantum mechanical operators acting on the Fock space states, with basic one-particle states conveniently labelled by eigenvalues of the momentum and spin operators.

In addition to continuous symmetries, whose infinitesimal generators (conserved Noether charges) form the Poincar\'e algebra, in quantum field theory we have to do with discrete symmetries: charge conjugation C, parity P, and time reversal T. Although individual discrete symmetries might be violated in some of the physical processes, their composition CPT is believed to be an exact symmetry of the standard relativistic quantum field theory.

When gravitational effects are taken into account, global spacetime symmetries are not symmetries of quantum field theory anymore. As a result of the quantum uncertainties expected in quantum gravity,  the   quantum spacetime ceases to be smooth and is replaced by spacetime foam, representing a random structure with no global spacetime symmetries at all. One can speculate however, that there exists an intermediate regime\footnote{This semiclassical regime is of major interest from the quantum gravity phenomenology perspective, see Refs.\cite{AmelinoCamelia:2008qg,AmelinoCamelia:2003ex,Addazi:2021xuf}.}, in which spacetime symmetries are still present, but have a form of deformation of the Poincar\'e symmetry. The parameter of this deformation should vanish when gravity is being switched off and it is customary to use in this role the inverse of Planck mass $1/\kappa$. Since this parameter has a dimension of length it can be identified with the structure constant of a Lie type spacetime noncommutativity induced by quantum gravity \cite{Doplicher:1994tu}. In this paper we are interested in a particular type of noncommutative spacetime, called $\kappa$-Minkowski space \cite{Lukierski:1993wx,Majid:1994cy} with the noncommutativity of the $\sf{AN}(3)$ type
\begin{equation}\label{0.1}
 [\hat x^0, \hat x^j] = \frac{i}{\kappa}\,  \hat x^j\,,\quad [\hat x^i, \hat x^j] =0\,.
\end{equation}
Alternatively, the square of the parameter $1/\kappa$ can be regarded as a curvature of the momentum space \cite{Majid:1999tc}, being in one-to-one correspondence with spacetime noncommutativity. In the case of Eq. \eqref{0.1} we have to do with momentum space having the form of the de Sitter space \cite{KowalskiGlikman:2002ft,KowalskiGlikman:2003we} embedded in the five-dimensional Minkowski space where
\begin{equation}\label{II.1.21}
   -p_0^2 + \mathbf{p}^2 + p_4^2 =\kappa^2\,.
\end{equation}
The symmetries of the noncommutative spacetime \eqref{0.1} and/or momentum space \eqref{II.1.21} form  a Hopf deformation of the Poincar\'e algebra, well-known under the name of the $\kappa$-Poincar\'e algebra \cite{Lukierski:1991pn,Lukierski:1992dt,Lukierski:1993wx,Majid:1994cy} (see the recent monograph \cite{Arzano:2021scz} for a comprehensive review.) The investigations of a free quantum field theory and its symmetries defined on the $\kappa$-Minkowski spacetime/de Sitter momentum space are the subject of the present paper.

In the recent paper \cite{Arzano:2020jro} we constructed a new theoretical setup for investigations of free complex quantum scalar field defined on the $\kappa$-Minkowski space. The major novelty of this approach was the introduction of two copies of the momentum space, associated with particles and antiparticles states. Together with the use of the antipode in the definition of the fields (see eq. \eqref{fa} and \eqref{fb} below), this in turn made it possible for us to obtain a well-defined and simple action of the discrete transformations C,P,T on the field. It was then argued that there is a subtle difference between Lorentz transformations of particles and antiparticles, which results in deformation of charge conjugation operator ${\cal C}$. Due to the complexity of the direct formulae, in \cite{Arzano:2020jro} we only computed the translational charges, stopping short of computing the boost ones. However, to fully appreciate the model and its features, a full formal treatment of all the charges was needed. It is the aim of this paper to compute boost charges of the model in \cite{Arzano:2020jro} and their brackets with the charge operator, so as to justify the claim about the deformation of  $\cal C$, and at the same time concluding the discussion about the symmetries of the model.

The plan of this paper is as follows. In the next section, we introduce some technical tools necessary for the definition of the deformed scalar field theory. Then we discuss the on-shell field and the associated symplectic structure. This presentation differs in some points from that presented in Ref. \cite{Arzano:2020jro}. Then, in Section III, we introduce the covariant phase space formalism, which gives a very simple and straightforward way of computing  Noether charges. Using it, we construct the conserved charges associated with  $\kappa$-Poincar\'e symmetry and compute their algebra. These charges are of primary importance here because, after quantization, the action of the associated operators  tells us how the symmetry acts on the states of the theory. We also notice a surprising property of the action of boosts on spacetime fields:  the deformed boost is here accompanied by a non-local spacetime  translation, different for positive and negative energy states. In the following section we consider all three discrete, deformed symmetries and explicitly construct the deformed charge-conjugation operator ${\cal C}$. We show a surprising fact that boost operators do not commute with ${\cal C}$ \footnote{Neither do the translations P, and we will discuss possible phenomenological consequences of this in a forthcoming paper.} and we discuss the physical consequences of this fact in details.  In Section V we revisit the relation between Lorentz symmetry and CPT in undeformed quantum field theories and we argue that the famous Jost-Wightman-Greenberg theorem \cite{Greenberg:2002uu} does not apply to the $\kappa$-deformed case. In the last Section VI we  describe possible phenomenological consequences of the results obtained in the paper, in particular on the lifetimes of particles and antiparticles. We finish the paper with conclusions.

\section{Preliminaries}\label{SectII}

In this section we will set up the model that we are going to consider in this paper. It differs in some details from the model discussed in Ref.  \cite{Arzano:2020jro}, but it shares with it all the basic properties. In particular, thanks to the field decomposition in eq. \eqref{fa}, \eqref{fb} introduced below, and thanks to the choice of the action \eqref{actionst}, the charge conjugation C can be defined in a particularly simple way.  The reader can find more details and a general background in the recent monograph \cite{Arzano:2021scz}.

We will mainly deal with the momentum-space picture of the $\kappa$-Poincar\'e algebra. The starting point is given by an explicit matrix representation of the algebra  \eqref{0.1}
\begin{equation}\label{II.1}
	\hat x^0 = -\frac{i}{\kappa} \,\left(\begin{array}{ccc}
		0 & \mathbf{0} & 1 \\
		\mathbf{0} & \mathbf{0} & \mathbf{0} \\
		1 & \mathbf{0} & 0
	\end{array}\right), \quad
	\hat{\mathbf{x}} = \frac{i}{\kappa} \,\left(\begin{array}{ccc}
		0 & {\mathbf{\epsilon}\,{}^T} &  0\\
		\mathbf{\epsilon} & \mathbf{0} & \mathbf{\epsilon} \\
		0 & -\mathbf{\epsilon}\,{}^T & 0
	\end{array}\right).
\end{equation}
%Notice that the central $\bm{0}$ in the above matrices is a null $3\times 3$ matrix, and the $\epsilon$ are column unitary vectors of three components. In particular, to $\hat{x}^1$ corresponds $\epsilon^T = (1,0,0)$, and analogous for the other components. 

From these matrices one can build the group elements of $\sf{AN}(3)$, defined by\footnote{We choose the time-to-the-right ordering for the product in the definition of $\hat{e}_k$. In this way we just choose particular coordinates in momentum space with no physical consequences \cite{Arzano:2021scz}. }
\begin{align}\label{II.2}
    \hat  e_k =e^{ik_i \hat x^i} e^{ik_0 \hat x^0}
\end{align}
which can be explicitly written as 
\begin{equation}\label{II.3}
 \hat  e_k   =\left(\begin{array}{ccc}
 \frac{ \bar p_4}\kappa \;&\;  \frac{\mathbf k}\kappa \; &\;
\frac{  p_0}\kappa\\&&\\
  \frac{\mathbf p}\kappa  & \mathbf{1} & \frac{\mathbf p}\kappa  \\&&\\
  \frac{\bar p_0}\kappa\; & -\frac{\mathbf k}\kappa\; &
\frac{  p_4}\kappa
\end{array}\right)\, 
\end{equation}
where 
\begin{eqnarray}\label{II.4}
 {p_0}(k_0, \mathbf{k}) &=&\kappa  \sinh
\frac{k_0}{\kappa} + \frac{\mathbf{k}^2}{2\kappa}\,
e^{  {k_0}/\kappa}, \nonumber\\
 p_i(k_0, \mathbf{k}) &=&   k_i \, e^{  {k_0}/\kappa},\label{II.1.20}\\
 {p_4}(k_0, \mathbf{k}) &=& \kappa \cosh
\frac{k_0}{\kappa} - \frac{\mathbf{k}^2}{2\kappa}\, e^{
{k_0}/\kappa}\nonumber
\end{eqnarray}
and
\begin{eqnarray}\label{6.1}
{\bar p_0}(k_0, \mathbf{k}) = \kappa  \sinh
\frac{k_0}{\kappa} - \frac{\mathbf{k}^2}{2\kappa}\,
e^{{k_0}/\kappa}, \nonumber \\ {\bar p_4}(k_0, \mathbf{k}) = \kappa \cosh
\frac{k_0}{\kappa} + \frac{\mathbf{k}^2}{2\kappa}\, e^{{k_0}/\kappa}.
\end{eqnarray}
It is easy to see that $(p_0, p_i, p_4)$ (which define what is called the classical basis) satisfy eq. \eqref{II.1.21}, and they can be interpreted as coordinates on (half of) the de Sitter momentum space defined by the conditions
\begin{align}
    p_+=p_0+p_4>0,
    \qquad
    p_4>0.
\end{align}
Notice that, given a momentum space origin $\mathcal{O} = (0,\dots,0,\kappa)^T$, one has 
\begin{align}
    (p_0, p_i, p_4)^T = \hat{e}_k \mathcal{O}.
\end{align}

\noindent We also need to introduce the antipodal map $S(p)$ with group element $\hat{e}_k$ defined in terms of the inverse of the $\sf{AN}(3)$ group element $\hat{e}_k$ as follows
\begin{align}
    (S(p)_0, S(p)_i, S(p)_4)^T = \hat{e}^{-1}_k \mathcal{O}
\end{align}
and having the form
\begin{equation}\label{II.5}
  S(p_0) = -p_0 + \frac{\mathbf{p}^2}{p_0+p_4} = \frac{\kappa^2}{p_0+p_4}-p_4\,,\quad S(\mathbf{p}) =-\frac{\kappa \mathbf{p} }{p_0+p_4}\,,\quad S(p_4) = p_4.
\end{equation}

The sum of momenta is defined through the group product of two $\sf{AN}(3)$ elements as follows
\begin{align}
    \hat{e}_k \hat{e}_l = \hat{e}_{k\oplus l}.
\end{align}
In terms of the classical basis \eqref{II.4}, this translates to
\begin{align}
(p\oplus q)_0 &= \frac1\kappa\, p_0(q_0+q_4) + \frac{\mathbf{p}\mathbf{q}}{p_0+p_4} +\frac{\kappa}{p_0+p_4}\, q_0\nonumber\\
(p\oplus q)_i &=\frac1\kappa\, p_i(q_0+q_4) + q_i\nonumber\\
(p\oplus q)_4 &= \frac1\kappa\, p_4(q_0+q_4) - \frac{\mathbf{p}\mathbf{q}}{p_0+p_4} -\frac{\kappa}{p_0+p_4}\, q_0\label{II.6}
\end{align}
with the property that $S(p)\oplus p = p \oplus S(p) = 0$. Notice that the above deformed summation rules are valid also if we exchange one or both of the momenta for their dual counterparts. 

The composition of momenta is directly related to the co-product $\Delta$, which dictates how deformed $\kappa$-Poincar\'e algebra generators of time translation (energy) $P_0$, space translation (linear momenta) $P_i$, boosts $N_i$ and rotations $R_i$ act on products of functions. We have
\begin{itemize}
    \item Translations
    \begin{align}
\Delta P_i &= \frac1\kappa\, P_{i}\otimes P_+ +\bbbone\otimes P_{i},\label{Ti}\\
\Delta P_0 &= \frac1\kappa\, P_{0}\otimes P_++\sum P_{k}P_+^{-1}\otimes P_{k}
 +\kappa\,P_{+}^{-1}\otimes P_{0}, \label{T0}\\
 \Delta P_4 &= \frac1\kappa\, P_{4}\otimes P_+-\sum P_{k}P_+^{-1}\otimes P_{k}
 -\kappa\,P_{+}^{-1}\otimes P_{0}, \label{T4}
\end{align}
where $P_+\equiv P_0 + P_4$, and for completeness we added the co-product for the operator $P_4$;
\item Boosts
    \begin{align}
\Delta N_i &=  N_{i}\otimes  \bbbone+\kappa\,P_+^{-1} \otimes N_{i} + \epsilon_{ijk} P_+^{-1}\, P_j \otimes M_k,
\label{Ni}
\end{align}
\item Rotations
   \begin{align}
\Delta M_i &=  M_{i}\otimes  \bbbone+  \bbbone\otimes M_{i}. 
\label{Mi}
\end{align}
\end{itemize}

In order to simplify the discussion, it will be convenient to go from the group elements $\hat{e}_k$ to canonical plane waves in commuting spacetime but with deformed sum of momenta given by eq. \eqref{II.6}. One can achieve this through the Weyl map $\mathcal{W}$ in the following way\footnote{There Weyl map is not unique, and therefore one can choose it in several ways. The approach taken in our previous work and in the present one is to use a Weyl map  identified by a well-defined action on derivatives. For more details, see Ref. \cite{Arzano:2020jro}. }
\begin{align}
    \mathcal{W}(\hat{e}_k \hat{e}_l)
    \equiv
    e_{p(k)} \star e_{q(l)}
    =
    e_{p\oplus q},
\end{align}
where now 
\begin{equation}
  e_p(x) = e^{ip_\mu\, x^\mu} = e^{-i(\omega_\mathbf{p}t-\mathbf{p}\mathbf{x})}.
\end{equation}

With these tools at hands, we can now introduce the action of the free scalar field, which is the most general  real action, bilinear in the fields and spacetime derivatives\footnote{See e.g., Refs. \cite{Freidel:2007hk} and  \cite{Arzano:2020jro} for definition of star product, deformed derivatives and their properties. In this paper, we use the mostly plus metric convention.}
\begin{equation}\label{actionst}
  S = -\frac{1}{2} \int_{ \mathbb{R}^4}d^4x\,
  \left[ (\partial_\mu \phi)^\dag\star\partial^\mu \phi
  + (\partial_\mu \phi) \star(\partial^\mu \phi)^\dag
  + m^2 (\phi^\dag\star \phi + \phi \star \phi^\dag)\right].
\end{equation}

Let us stop here to comment on the form of this action. The first thing to notice is that it is  highly non-local. As shown explicitly in Ref. \cite{Kowalski-Glikman:2009len} the spacetime representation of star product is an operator of the form $\sum_n (\partial/\kappa)^n$. It turns out, however  that the field equations are still the same, as it is in the case of the undeformed field theory. Therefore the actions of free $\kappa$-deformed and undeformed fields differ, essentially, by boundary terms. This might seem innocent at the first sight, but in fact the difference is essential. The boundary terms do not change the bulk equations of motion but they do influence the form of symplectic structure and of the Noether charges. Since the latter are measurable, they are directly related physical symmetries. In the $\kappa$-deformed case, where the invariances of the action and of the symplectic form do not coincide, it is the charges and their algebra that define the symmetries of the theory.

Following Ref.  \cite{Arzano:2020jro}, we define the on-shell field as follows
\begin{align}\label{fa}
	\phi(x) &= \int \frac{d^3p}{\sqrt{2\omega_\mathbf{p}}}\, \,\zeta(p)\, a_{\mathbf{p}}\, e^{-i(\omega_\mathbf{p}t-\mathbf{p}\mathbf{x})}  + \int \frac{d^3p}{\sqrt{2\omega_\mathbf{p}}}\, \,\zeta(p) b^\dag_{\mathbf{p}}\, e^{-i(S(\omega_\mathbf{p})t-S(\mathbf{p})\mathbf{x})}, 
\end{align}
\begin{align}\label{fb}
	\phi^\dag(x) &= \int \frac{d^3p}{\sqrt{2\omega_\mathbf{p}}}\, \,\zeta(p)\, a^\dag_{\mathbf{p}}\, e^{-i(S(\omega_\mathbf{p})t-S(\mathbf{p})\mathbf{x})}  + \int \frac{d^3p}{\sqrt{2\omega_\mathbf{p}}}\, \,\zeta(p) b_{\mathbf{p}}\, e^{-i(\omega_\mathbf{p}t-\mathbf{p})\mathbf{x}}, 
\end{align}
where $\zeta(p)$ is at this stage an arbitrary function which will be fixed in below. 

Notice some particular features of the field definition. The first is that contrary to the choice made in the most of the field theoretical constructions in the past in \cite{Arzano:2020jro} and here we choose to separate positive and negative energy states and manifestly label the negative energy plane waves with the antipode. This simplifies the treatment of the negative energy states making it also possible to manifestly avoid the problem of Lorentz symmetry violation (first noticed in \cite{Bruno:2001mw}, discussed in depth in \cite{Freidel:2007hk} and finally resolved in \cite{Arzano:2009ci}) Furthermore, it is the presence of the antipode in the negative energy plane waves (and its absence in the positive energy ones) which is responsible for the simple definition of charge conjugation in this model.

Substituting Eqs. \eqref{fa} and \eqref{fb} to the action \eqref{actionst} we obtain the momentum space action of the form
$$
	S
	=
	\frac{1}{2}
	\int\frac{d^3p}{{2\omega_\mathbf{p}}}\, 
	\zeta(p)^2
	\left(
	1 + \frac{|p_+|^3}{\kappa^3}
	\right)
	\left[
	(p_\mu p^\mu + m^2)
	 a_{\mathbf{p}}a_{\mathbf{p}}^\dag
	+
	(S(p)_\mu S(p)^\mu + m^2)
	b_{\mathbf{p}} b_{\mathbf{p}}^\dag
	\right]
$$
which by noticing that $S(p)_\mu S(p)^\mu = p_\mu p^\mu$ simplifies to give
\begin{align}\label{actionms}
	S
	=
	\frac{1}{2}
	\int\frac{d^3p}{{2\omega_\mathbf{p}}}\, 
	\zeta(p)^2
	\left(
	1 + \frac{|p_+|^3}{\kappa^3}
	\right)\left(p_\mu p^\mu + m^2\right)
	\left[
	a_{\mathbf{p}}^\dag a_{\mathbf{p}} 
	+
	b_{\mathbf{p}}^\dag b_{\mathbf{p}}
	\right]
\end{align}
and we see that the on-shell action is identically equal to zero. Furthermore, from the action one can also obtain the symplectic form, which reads \cite{Arzano:2020jro}
\begin{align}\label{Omega}
	\Omega 
	&=
-	2i\int \frac{d^3p}{2\omega_p}
	\zeta(p)^2	\left[
	\frac{|p_+|^3}{\kappa^3}
	\omega_p - S(\omega_p)
	\right]
	\frac{p_4}{\kappa}\left(
	\delta a^\dag_\mathbf{p}\wedge
	\delta a_\mathbf{p}\,
-
	\delta b_{\mathbf{p}}\wedge
	\delta b_{\mathbf{p}}^\dag
	\right) 
 \nonumber\\
&\equiv- i\int {d^3p}\,\alpha(p)
	\left(	\delta a^\dag_\mathbf{p}\wedge
	\delta a_\mathbf{p}\, - \delta b_{\mathbf{p}}\wedge
	\delta b^\dag_{\mathbf{p}}\right) 
\end{align}
where we introduced a new coefficient function
\begin{align}\label{alpha}
\alpha(p) = \frac{1}{2\omega_p}\,	\zeta(p)^2 \frac{p_4}{\kappa}\, \left[
	\frac{|p_+|^3}{\kappa^3}
	\omega_p - S(\omega_p)
	\right].	
\end{align}
Notice that it is not possible to make the action and the symplectic form simultaneously Lorentz invariant, and we choose to fix the parameter 
 $\zeta(p)$ to be
\begin{align}\label{convention}
    \zeta(p)^{-2} = \frac{1}{2\omega_p}\, \frac{p_4}{\kappa}\, \left[
	\frac{|p_+|^3}{\kappa^3}
	\omega_p - S(\omega_p)
	\right],
\end{align}
resulting in $\alpha(p)=1$ and the simplest possible symplectic form (and as a matter of fact, since $\alpha(p)\rightarrow 1$ in the limit $\kappa\rightarrow\infty$, this simple symplectic form is formally the same as the undeformed one). We will work in this convention in the remainder of the paper.

It follows immediately from this expression that the commutation relations for the creation/annihilation operators are
\begin{align}\label{aadagalg}
    [\hat a_\mathbf{p},\hat a^\dag_\mathbf{q}] &= \delta^3(\mathbf{p}-\mathbf{q}),\nonumber\\
     [\hat b_\mathbf{p},\hat b^\dag_\mathbf{q}] &=\delta^3(\mathbf{p}-\mathbf{q}).
\end{align}

Let us pause here for a moment to make an important comment. The  construction of the creation/annihilation operators algebra and of the Fock space in deformed theories was a long- standing problem \cite{Arzano:2007ef,Arzano:2007nx,Daszkiewicz:2007ru,Arzano:2013sta}  which was solved only recently and the solution will be presented in the forthcoming paper \cite{ArzanoKowalski}. In this paper we do not go beyond the one-particle states and therefore all the subtleties of the deformed Fock space construction are not  relevant for our present discussion.

\section{Conserved charges: geometric approach}

In this section we derive the form of conserved charges associated with deformed Poincar\'e symmetry. We know from Noether theorem that there is the one-to-one correspondence between the symmetries and charges conservation. In what follows we  define the symmetries  by demanding conservation of the associated charges . Then we show that Poisson brackets of the charges we constructed form a representation of the Poincar\'e algebra.

In order to compute the charges associated with symmetries of the action we make use  of the covariant phase space formalism (cf. Ref.\ \cite{Harlow:2019yfa} for a comprehensive recent review and references to the original works and \cite{Arzano:2007gr} for a past attempt to used this formalism in the context of $\kappa$-deformed theories.). This approach allows us a much more direct computation of the charges, which were otherwise proved to be prohibitively difficult, starting from the action and taking its variation. The conserved charge  $Q_\xi$ related to the symmetry associated with the continuous vector field (generator of an infinitesimal transformation) $\xi$  is defined by the equation
\begin{align}\label{chargexi}
	-\delta_\xi \lrcorner \, \Omega  = \delta Q_\xi
\end{align}
where $\delta$ is the exterior derivative in the phase space, i.e.\ on the set of solutions of the equations of motion. The object $\delta_\xi$ is a vector field in phase space generated by the vector field $\xi$. In other words, $\delta_\xi A$ measures the infinitesimal variation of the object $A$ in phase space due to the action of the symmetry transformation along $\xi$ in spacetime. The symbol $\lrcorner$ denotes contraction of the vector field with the form.

\subsection{Undeformed case}

Let us first  compute the charges in  case of the standard, undeformed field theory. The symplectic form $\Omega$ is given by eq.\ \eqref{Omega} above and for $\delta_\xi$ we use the standard transformation rules of the Fourier components of the field.

\subsubsection{Translation charges}

Starting with the time translations, we first need to define the variation of the field components $\delta^T_{0}a_\mathbf{p}$, $\delta^T_{0}b_\mathbf{p}$, where by $\delta^T_{0}$ we denote the translation generated by an infinitesimal parameter $\epsilon^\mu=(\epsilon, \mathbf{0})$. We know translations act on the Fourier components as phase transformations
\begin{align}
	a_\mathbf{p} \mapsto e^{i\epsilon \omega_p}a_\mathbf{p}
	\approx
	a_\mathbf{p}
	+
	i\epsilon \omega_p a_\mathbf{p}
\end{align}
and therefore
\begin{align}\label{undeftrans}
	\delta^T_{0} a_\mathbf{p} = i\epsilon \omega_p a_\mathbf{p}
	\quad
	\Leftrightarrow
	\quad
	\delta^T_{0} a_\mathbf{p}^\dag
	=
	-
	i\epsilon \omega_p a_\mathbf{p}^\dag,
\end{align}
where the second expression can be obtained from the first one by  taking the Hermitian conjugate. Therefore, we have 
\begin{align}
	-\delta^T_{0} \lrcorner \, \Omega
	&=
	i\int d^3p \,
	(
	\delta^T_{0} a^\dag_\mathbf{p}
	\delta a_\mathbf{p}\,
	-
	\delta a^\dag_\mathbf{p}
	\delta^T_{0} a_\mathbf{p}
	-
\delta^T_{0} b_{\mathbf{p}}
	\delta b^\dag_{\mathbf{p}}
	+
	\delta b_{\mathbf{p}}
	\delta^T_{0} b^\dag_{\mathbf{p}})\nonumber \\
	&=
	\epsilon
	\delta
	\int d^3p \,
	\omega_p
	(
	a_\mathbf{p}
	a_\mathbf{p}^\dag\,
	+
	b_{\mathbf{p}}^\dag
	b_{\mathbf{p}}
	)\label{eq32}
\end{align}
and we get the canonical time-translational charge, the Hamiltonian
$$
{\cal H}^{\kappa \rightarrow \infty} = 	\int d^3p \,
	\omega_p
	(
	a_\mathbf{p}^\dag\,a_\mathbf{p}
	+
	b_{\mathbf{p}}^\dag
	b_{\mathbf{p}}
	)
$$
multiplied by the infinitesimal parameter $\epsilon$. It should be stressed that in the course of  computing $-\delta^T_{0} \lrcorner \, \Omega$ we made use of the standard Leibniz rule. This is the aspect of this calculation that is going to change when we turn to the deformed case. Also, the fact that $-\delta^T_{0} \lrcorner \, \Omega$ turned out to be a total differential $\delta {\cal H}^{\kappa \rightarrow \infty}$ confirms that the phase transformation is a symmetry of the free field (in momentum space representation).

The same exact reasoning works for the spatial translation charges, where it is sufficient to replace $\omega_p$ with ${p}_i$ everywhere, so that
\begin{align}
	-\delta^T_{i} \lrcorner \, \Omega
	&=
	\epsilon^i
	\delta
	\int d^3p \,
	p_i
	(
	a_\mathbf{p}^\dag\,a_\mathbf{p}
	+
	b_{\mathbf{p}}^\dag
	b_{\mathbf{p}}
	)
\end{align}
and the corresponding charge is equal to
$$
{\cal P}_i^{\kappa \rightarrow \infty} = 	\int d^3p \,
	p_i
	(
	a_\mathbf{p}^\dag
	a_\mathbf{p}
	+
	b_{\mathbf{p}}^\dag
	b_{\mathbf{p}}
	).
$$
Notice that one can easily get how the field transforms under the transformation \eqref{undeftrans} (and the equivalent one for the spatial translation). One has 
\begin{align}\label{ftrans}
\delta^T	\phi(x) 
&=  
\int \frac{d^3p}{\sqrt{2\omega_\mathbf{p}}}\, \,\zeta(\mathbf{p})\, i \epsilon^\mu p_\mu\, a_{\mathbf{p}}\, e^{-i(\omega_\mathbf{p}t-\mathbf{p}\mathbf{x})}  + \int \frac{d^3p}{\sqrt{2\omega_\mathbf{p}}}\, \,\zeta(\mathbf{p})\,  i \epsilon^\mu S(p)_\mu b^\dag_{\mathbf{p}}\, e^{-i(S(\omega_\mathbf{p})t-S(\mathbf{p})\mathbf{x})}\nonumber \\
&\equiv  
\epsilon^\mu\partial_\mu\, \phi(x)
\end{align}
with $p_\mu = (\omega_\mathbf{p},\mathbf{p} )$, which proves that the infinitesimal spacetime translation indeed corresponds to phase transformation of Fourier components.

\subsubsection{Lorentz charges: boosts and rotations}

In the case of boosts the field Fourier components transform as follows
\begin{align}\label{assumedboost}
	\delta^B a_\mathbf{p} &=i\omega_{\mathbf p}\, \lambda^j
	\frac{\partial a_\mathbf{p}}{\partial \mathbf{p}^j}\,
	+
	i\, a_\mathbf{p}
	\lambda^j
	\frac{\mathbf{p}_j}{2\omega_\mathbf{p}}
	,\quad \nonumber \\ 
	\delta^B a^\dag_\mathbf{p} &=i\omega_{\mathbf p}\, \lambda^j
	\frac{\partial a^\dag_\mathbf{p}}{\partial \mathbf{p}^j}
	+
	i\, a_\mathbf{p}^\dag
	\lambda^j
	\frac{\mathbf{p}_j}{2\omega_\mathbf{p}}
\end{align}
and analogous expressions for $b_\mathbf{p}$, $b^\dag_\mathbf{p}$.
 Then we have
\begin{align}\label{boostcomp}
-	\delta^B \lrcorner \, \Omega^{\kappa \rightarrow \infty}
	&=
i	\int d^3p \,
	(
	\delta^B a^\dag_\mathbf{p}
	\delta a_\mathbf{p}\,
	-
	\delta a_\mathbf{p}^\dag
	\delta^B a_\mathbf{p}
	-
	\delta^B b_{\mathbf{p}}
	\delta b^\dag_{\mathbf{p}}
	+
	\delta b_{\mathbf{p}}
	\delta^B b^\dag_{\mathbf{p}}) \nonumber\\
%%%%%%%%%%%%%%%%% second step
	&=
    {\lambda}^i \int d^3p \,
	\omega_{\mathbf p}
	\left(
	\frac{\partial a_\mathbf{p}}{\partial \mathbf{p}^i}
	\delta a_\mathbf{p}^\dag\,
	-
	\delta a_\mathbf{p}
	\frac{\partial a_\mathbf{p}^\dag}{\partial \mathbf{p}^i}
	-
	\frac{\partial b_{\mathbf{p}}^\dag}{\partial \mathbf{p}^i}
	\delta b_{\mathbf{p}}
	+
	\delta b_{\mathbf{p}}^\dag
	\frac{\partial b_{\mathbf{p}}}{\partial \mathbf{p}^i}
	\right) \nonumber \\
	&+
	{\lambda}^i \int d^3p \,
	\frac{\mathbf{p}_i}{2\omega_\mathbf{p}}
    \left(
	a_\mathbf{p}
	\delta a_\mathbf{p}^\dag\,
	-
	\delta a_\mathbf{p}
	a_\mathbf{p}^\dag
	+
	\delta b_\mathbf{p}^\dag\,
	b_\mathbf{p}
	-
	b_\mathbf{p}^\dag
	\delta b_\mathbf{p}
	\right) \nonumber \\
%%%%%%%%%%%%%%%%%%%%%% third step
	&=
	\frac{1}{2}
    {\lambda}^i \delta \int d^3p \,
	\omega_{\mathbf p}
	\left(
	\frac{\partial a_\mathbf{p}}{\partial \mathbf{p}^i}
	a_\mathbf{p}^\dag\,
	-
	a_\mathbf{p}
	\frac{\partial a_\mathbf{p}^\dag}{\partial \mathbf{p}^i}
	-
	\frac{\partial b_{\mathbf{p}}^\dag}{\partial \mathbf{p}^i}
	b_{\mathbf{p}}
	+
	b_{\mathbf{p}}^\dag
	\frac{\partial b_{\mathbf{p}}}{\partial \mathbf{p}^i}
	\right)
\end{align}
which gives the undeformed boost charge
\begin{align}
    {\cal N}_i^{\kappa \rightarrow \infty} =
    \frac{1}{2}
 \int d^3p \,
	\omega_{\mathbf p}
	\left(
	\frac{\partial a_\mathbf{p}}{\partial \mathbf{p}^i}
	a_\mathbf{p}^\dag\,
	-
	a_\mathbf{p}
	\frac{\partial a_\mathbf{p}^\dag}{\partial \mathbf{p}^i}
	-
	\frac{\partial b_{\mathbf{p}}^\dag}{\partial \mathbf{p}^i}
	b_{\mathbf{p}}
	+
	b_{\mathbf{p}}^\dag
	\frac{\partial b_{\mathbf{p}}}{\partial \mathbf{p}^i}
	\right).
\end{align}
Notice that in the last passage of Eq. \eqref{boostcomp} we used the fact that (we consider only the $a, a^\dag$ case since the same reasoning applies to $b, b^\dag$)
\begin{align}
	\frac{\partial a_\mathbf{p}}{\partial \mathbf{p}^i}
	\delta a_\mathbf{p}^\dag\,
	-
	\delta a_\mathbf{p}
	\frac{\partial a_\mathbf{p}^\dag}{\partial \mathbf{p}^i}
	=
	\frac{1}{2}
	\delta
	\left(
	\frac{\partial a_\mathbf{p}}{\partial \mathbf{p}^i}
	a_\mathbf{p}^\dag\,
	-
	a_\mathbf{p}
	\frac{\partial a_\mathbf{p}^\dag}{\partial \mathbf{p}^i}
	\right) 
	+
	\frac{1}{2}
	\frac{\partial }{\partial \mathbf{p}^i}
	\left(
	a_\mathbf{p}
	\delta a_\mathbf{p}^\dag\,
	-
	\delta a_\mathbf{p}
	a_\mathbf{p}^\dag
	\right).
\end{align}
The second term on the RHS of the last equation is still not a total derivative because of the presence of $\omega_\mathbf{p}$ inside the integrand.
After integration by parts, this term contributes a factor 
\begin{align}
    -\frac{\mathbf{p}_i}{2\omega_\mathbf{p}}
    \left(
	a_\mathbf{p}
	\delta a_\mathbf{p}^\dag\,
	-
	\delta a_\mathbf{p}
	a_\mathbf{p}^\dag
	\right)
\end{align}
to the integrand, which cancels with the second term in the second passage of Eq. \eqref{boostcomp}.

The case of rotations $R$ is analogous, and starting from the transformations
\begin{align}\label{dlorentzu}
	\delta^R a_\mathbf{p}
	&=
	i \epsilon^{ijk} \rho_i
	\mathbf{p}_{j} \frac{\partial}{\partial \mathbf{p}^{k}}
	a_\mathbf{p} 
\qquad
	\delta^R a_\mathbf{p}^\dag
	=
	i \epsilon^{ijk} \rho_i
	\mathbf{p}_{j} \frac{\partial}{\partial \mathbf{p}^{k}}
	a_\mathbf{p}^\dag 
\end{align}
and analogously for $b_\mathbf{p}, b^\dag_\mathbf{p}$, one gets the charge
\begin{align}
    {\cal M}_i =
    \frac{1}{8}
 \int d^3p \,\epsilon_{ijk}
	\mathbf{p}^j
	\left(
	\frac{\partial a_\mathbf{p}}{\partial \mathbf{p}_k}
	a_\mathbf{p}^\dag\,
	-
	a_\mathbf{p}
	\frac{\partial a_\mathbf{p}^\dag}{\partial \mathbf{p}_k}
	-
	\frac{\partial b_{\mathbf{p}}^\dag}{\partial \mathbf{p}_k}
	b_{\mathbf{p}}
	+
	b_{\mathbf{p}}^\dag
	\frac{\partial b_{\mathbf{p}}}{\partial \mathbf{p}_k}
	\right).
\end{align}

Once again, we can obtain the field transformations from Eqs. \eqref{assumedboost} and \eqref{dlorentzu} , respectively
\begin{align}
\delta^B\phi(x) 
& =
i\lambda^j\, 
\int \frac{d^3p}{\sqrt{2\omega_\mathbf{p}}}\, \,
\left[
\omega_{\mathbf p}\, 
	\frac{\partial a_\mathbf{p}}{\partial \mathbf{p}^j}\,
	+
	a_\mathbf{p}
	\frac{\mathbf{p}_j}{2\omega_\mathbf{p}}
\right]
e^{-i(\omega_\mathbf{p}t-\mathbf{p}\mathbf{x})}  
+
\left[
\omega_{\mathbf p}\, 
	\frac{\partial b^\dag_\mathbf{p}}{\partial \mathbf{p}^j}
	+
	b_\mathbf{p}^\dag
	\frac{\mathbf{p}_j}{2\omega_\mathbf{p}}
\right]
e^{i(\omega_\mathbf{p}t-\mathbf{p}\mathbf{x})} \nonumber \\
&=
-i\lambda_j\, 
\int \frac{d^3p}{\sqrt{2\omega_\mathbf{p}}}\, 
a_{\mathbf{p}}\,
\omega_\mathbf{p} \frac{\partial}{\partial \mathbf{p}_j}
e^{-i(\omega_\mathbf{p}t-\mathbf{p}\mathbf{x})}  
+
b^\dag_{\mathbf{p}}\,
\omega_\mathbf{p} \frac{\partial}{\partial \mathbf{p}_j}\, 
e^{i(\omega_\mathbf{p}t-\mathbf{p}\mathbf{x})} \nonumber \\
&=i \lambda_j\, x^j\frac{\partial }{\partial t}\, \phi(x)
\end{align}
and 
\begin{align}
\delta^R \phi(x) 
&=
i\rho^i\, \epsilon_{ik}{}^j\, 
\int \frac{d^3p}{\sqrt{2\omega_\mathbf{p}}}\, 
a_{\mathbf{p}}\,
\mathbf{p}_k \frac{\partial}{\partial \mathbf{p}_j}
e^{-i(\omega_\mathbf{p}t-\mathbf{p}\mathbf{x})}  
+
i\rho^i\, \epsilon_{ik}{}^j\, 
\int \frac{d^3p}{\sqrt{2\omega_\mathbf{p}}}\, 
b^\dag_{\mathbf{p}}\,
\mathbf{p}_k \frac{\partial}{\partial \mathbf{p}_j}\, 
e^{i(\omega_\mathbf{p}t-\mathbf{p}\mathbf{x})} \nonumber\\
&=
i\rho^i\, \epsilon_{ik}{}^j\,  x^j\frac\partial{\partial x^k}\, \phi(x),
\end{align}
which are the standard spacetime boost and rotation transformations.

Having derived the undeformed charges, let us now turn to the deformed case.

\subsection{Deformed case}

Before we turn to detailed computations we must first understand what is the exact meaning of the formal expression $-\delta_\xi \lrcorner \, \Omega $ in the deformed case. We explain it by computing the simpler deformed translational charge and presenting all the details, and then we will consider Lorentz charges.

\subsubsection{Translational charges}

As we already have an explicit expression for the translation charges derived in Ref. \cite{Arzano:2020jro}, here we can describe  derivation of the same translation charges in the covariant phase space formalism. We postulate the following transformation rules  of the momentum space fields. Here, contrary to Ref. \cite{Arzano:2020jro}, we take as a starting point the momentum space action of the translation, not the spacetime ones:
\begin{align}
	\delta^T a_\mathbf{p} = i \epsilon^\mu p_\mu a_\mathbf{p},
	\qquad
	\delta^T a_\mathbf{p}^\dag = i \epsilon^\mu S(p)_\mu a_\mathbf{p}^\dag,
	\qquad
	\delta^T b_{\mathbf{p}}^\dag = i \epsilon^\mu S(p)_\mu b_{\mathbf{p}}^\dag,
	\qquad
	\delta^T b_{\mathbf{p}} = i \epsilon^\mu p_\mu b_{\mathbf{p}}.\label{deformedtransla}
\end{align}

Further, we define the deformed contraction of a vector field with the symplectic two-form
\begin{align}\label{deformedcontraction}
	\delta_\xi \lrcorner \, (\delta a^\dag_\mathbf{p}\wedge
	\delta a_\mathbf{p})
	&=
	(\delta_\xi a^\dag_\mathbf{p})
	\delta a_\mathbf{p}
	+
	\delta a^\dag_\mathbf{p} [S
	(
	\delta_\xi 
	)
	a_\mathbf{p} ], \nonumber \\
	\delta_\xi \lrcorner \, (\delta b_\mathbf{p}\wedge
	\delta b^\dag_\mathbf{p})
	&=
	(\delta_\xi b_\mathbf{p})
	\delta b^\dag_\mathbf{p}
	+
	\delta b_\mathbf{p} [S
	(
	\delta_\xi 
	)
	b^\dag_\mathbf{p} ].
\end{align}

Substituting these expressions into Eq.
\eqref{chargexi} we get
\begin{align}
	-\delta^T \lrcorner \, \Omega
	&=
	i\int d^3p \,
	(
	\delta^T a_\mathbf{p}^\dag
	\delta a_\mathbf{p}\,
	+
	\delta a_\mathbf{p}^\dag
	S(\delta^T) a_\mathbf{p}
	-
	\delta^T b_{\mathbf{p}}
	\delta b_{\mathbf{p}}^\dag
	-
	\delta b_{\mathbf{p}}
	S(\delta^T) b_{\mathbf{p}}^\dag)\nonumber \\
	&=
	-\epsilon^\mu \delta\left(\int d^3p \,
	[
	S(p)_\mu a_\mathbf{p}^\dag
	\delta a_\mathbf{p}\,
	-
	p_\mu b_{\mathbf{p}}^\dag
	\delta b_{\mathbf{p}}\right)
	\end{align}
and we find the translational charge to be
\begin{align}
    {\cal P}_\mu =\int d^3p \,
	[
	-S(p)_\mu a_\mathbf{p}^\dag
	 a_\mathbf{p}\,
	+
	p_\mu b_{\mathbf{p}}^\dag
	 b_{\mathbf{p}}
	 ],
\end{align}
in agreement with Ref. \cite{Arzano:2020jro}. 

It could be easily checked, following the steps analogous to that used in the undeformed case, that the deformed phase transformations \eqref{deformedtransla} lead to the same spacetime translations $\delta^T	\phi(x) =
\epsilon^\mu\partial_\mu\, \phi(x)$.

Let us pause for a moment to discuss important subtle points of the construction of deformed translational charge presented above.

The main difference between Eqs.\ \eqref{deformedcontraction} and the analogous relation in the undeformed context is the presence of the antipode instead of the  minus in the second terms. Notice that, since $S(AB)=S(A)S(B)$, and since $S(A)S(B) \xrightarrow{\kappa\rightarrow\infty}(-A)(-B) = AB$, the object $S(\delta_\xi)$ in Eq.\ \eqref{deformedcontraction} must contain an additional minus sign when necessary, for consistency with the undeformed axiom. For example, if $\delta_\xi = \epsilon \mathbf{p}_i \frac{\partial}{\partial \mathbf{p}_i}$ where $\epsilon$ is just some constant, then $S(\delta_\xi) = -\epsilon S(\mathbf{p})_i \frac{\partial}{\partial S(\mathbf{p})_i}$, which has the correct $\kappa\rightarrow\infty$ limit.

Secondly, in the derivation of the translational charges we made use of the symplectic form \eqref{Omega}, but one could ask what would happen if, for example, we use the expression
\begin{align}\label{Omega1}
	\Omega_1
	&=
- i\,\int {d^3p}\,
	\left(-	\delta a_\mathbf{p}\wedge
	\delta a^\dag_\mathbf{p}\, -\delta b_{\mathbf{p}}\wedge
	\delta b^\dag_{\mathbf{p}}\right) 
\end{align}
which would apparently lead to a different expression for the translational charges. The answer to this relies on the presence of the antipode in the contraction of a two-form with a vector field in Eq.\ \eqref{deformedcontraction}. In  the undeformed context, the contraction of a vector field with a general 2-form is related to the minus sign that one gets from exchanging the two 1-forms $w$ and $v$, since trivially 
\begin{align}
    &\delta_\xi \, \lrcorner \, (w\wedge v) 
    = 
    (\delta_\xi w) v - w (\delta_\xi v)= \nonumber \\
    &= -[
    w (\delta_\xi v) - (\delta_\xi w) v
    ]
    =
    - \delta_\xi \, \lrcorner \, (v\wedge w).
\end{align}
In particular, the minus sign that one gets from switching the order of the wedge product is `the same one' as the one in the definition of contraction of vector field with a 2-form. In the deformed context, this translates into 
\begin{align}
    \delta_\xi \, \lrcorner \, (w\wedge v) 
    =
    S(\delta_\xi) \, \lrcorner \, (v\wedge w)
\end{align}
which means that the treatment of the minus sign in presence of contractions is non-trivial.

\subsubsection{Lorentz charges}

Let us now consider the boost charges. Assuming the following transformations of the creation and annihilation operators 
\begin{align}\label{dboost}
	\delta^B a_\mathbf{p} 
	&=
	-i \lambda^j \, 
	\omega_\mathbf{p}
	\left[
	\frac{\partial}{\partial \mathbf{p}^j}
	+
	\frac{1}{2}
	\frac{1}{\omega_\mathbf{p}}
	\frac{\partial [\omega_\mathbf{p} ]}{\partial \mathbf{p}^j}
	\right]a_\mathbf{p}, \nonumber \\
	\delta^B a_\mathbf{p}^\dag
	&=
	-i \lambda^j \,
	S(\omega_\mathbf{p}) 
	\left[
    \frac{\partial }{\partial S(\mathbf{p})^j}
    +
    \frac{1}{2}
    \frac{1}{S(\omega_\mathbf{p})}
    \frac{\partial [S(\omega_\mathbf{p}) ]}{\partial S(\mathbf{p})^j}
    \right]a_{\mathbf{p}}^\dag, \nonumber \\
	\delta^B b_\mathbf{p} 
	&=
	-i \lambda^j \, 
	\omega_\mathbf{p}
	\left[
	\frac{\partial}{\partial \mathbf{p}^j}
	+
	\frac{1}{2}
	\frac{1}{\omega_\mathbf{p}}
	\frac{\partial [\omega_\mathbf{p} ]}{\partial \mathbf{p}^j}
	\right]b_\mathbf{p}, \nonumber \\
	\delta^B b_\mathbf{p}^\dag
	&=
	-i \lambda^j \,
	S(\omega_\mathbf{p}) 
	\left[
    \frac{\partial }{\partial S(\mathbf{p})^j}
    +
    \frac{1}{2}
    \frac{1}{S(\omega_\mathbf{p})}
    \frac{\partial [S(\omega_\mathbf{p}) ]}{\partial S(\mathbf{p})^j}
    \right]b_\mathbf{p}^\dag, 
\end{align}
we get
\begin{align}
	-&\delta^B \lrcorner \, \Omega
	=
-	i\int d^3p \,
	\left(
	\delta^B a_\mathbf{p}^\dag
	\delta a_\mathbf{p}\,
	+
	\delta a_\mathbf{p}^\dag
	S(\delta^B) a_\mathbf{p}
	-
	\delta^B b_{\mathbf{p}}
	\delta b_{\mathbf{p}}^\dag
	-
	\delta b_{\mathbf{p}}
	S(\delta^B) b_{\mathbf{p}}^\dag\right)  \nonumber\\
%%%%%%%%%%%%%%%%%%%%%%%%%%%%%%% FIRST STEP
	&=
-\lambda^i\,\int d^3p \,
	\left(
	 S(\omega_p) \frac{\partial a_\mathbf{p}^\dag}{\partial S(\mathbf{p})^i}
	\delta a_\mathbf{p}\,
	-
	\delta a_\mathbf{p}^\dag \,
	 S(\omega_p) \frac{\partial a_\mathbf{p}}{\partial S(\mathbf{p})^i}
	-
	 \omega_p\frac{\partial b_{\mathbf{p}}}{\partial \mathbf{p}^i}
	\delta b_{\mathbf{p}}^\dag
	+
	\delta b_{\mathbf{p}}
	 \omega_p \frac{\partial b_{\mathbf{p}}^\dag}{\partial \mathbf{p}^i}
	\right)\nonumber\\
%%%
	&
	+
	\lambda^i\,\int d^3p  \frac{1}{2}
	\frac{\partial [S(\omega_\mathbf{p}) ]}{\partial S(\mathbf{p})^i}
	\left(
	a_\mathbf{p}
	\delta a_\mathbf{p}^\dag\,
	-
	\delta a_\mathbf{p}
	a_\mathbf{p}^\dag
	\right)
	-
	\frac{1}{2}
	\frac{\partial [\omega_\mathbf{p} ]}{\partial \mathbf{p}^i}
	\left(
	b_{\mathbf{p}}
	\delta b_{\mathbf{p}}^\dag
	-
	\delta b_{\mathbf{p}} b_{\mathbf{p}}^\dag
	\right)  \\
%%%%%%%%%%%%%%%%%%%%%%%%%%%%%%%%%%%%%%% SECOND STEP
	&=
	-\frac{1}{2}{\lambda}^i \delta \int d^3p \,
	\left\{
	 S(\omega_p) \left[
	\frac{\partial a_\mathbf{p}^\dag}{\partial S(\mathbf{p})^i}
	a_\mathbf{p}\,
	-
	a_\mathbf{p}^\dag \, \frac{\partial a_\mathbf{p}}{\partial S(\mathbf{p})^i}
	\right]
	+
	\omega_p
	\left[
	b_{\mathbf{p}}
	\frac{\partial b_{\mathbf{p}}^\dag}{\partial \mathbf{p}^i}
	-
	\frac{\partial b_{\mathbf{p}}}{\partial \mathbf{p}^i}
	b_{\mathbf{p}}^\dag
	\right]
	\right\}
\end{align}
which gives the boost charge
\begin{align}\label{defboostfromsymp}
 {\cal   N }_i
    =
   - \frac{1}{2}
    \int d^3p \,
	\left\{
	 S(\omega_p) \left[
	\frac{\partial a_\mathbf{p}^\dag}{\partial S(\mathbf{p})^i}
	a_\mathbf{p}\,
	-
	a_\mathbf{p}^\dag \, \frac{\partial a_\mathbf{p}}{\partial S(\mathbf{p})^i}
	\right]
	+
	\omega_p
	\left[
	b_{\mathbf{p}}
	\frac{\partial b_{\mathbf{p}}^\dag}{\partial \mathbf{p}^i}
	-
	\frac{\partial b_{\mathbf{p}}}{\partial \mathbf{p}^i}
	b_{\mathbf{p}}^\dag
	\right]
	\right\}.
\end{align}
Once again, we used the relation
\begin{align*}
	\frac{\partial a_\mathbf{p}}{\partial S(\mathbf{p})^i}
	\delta a_\mathbf{p}^\dag\,
	-
	\delta a_\mathbf{p}
	\frac{\partial a_\mathbf{p}^\dag}{\partial S(\mathbf{p})^i}
	=
	\frac{1}{2}
	\delta
	\left(
	\frac{\partial a_\mathbf{p}}{\partial S(\mathbf{p})^i}
	a_\mathbf{p}^\dag\,
	-
	a_\mathbf{p}
	\frac{\partial a_\mathbf{p}^\dag}{\partial S(\mathbf{p})^i}
	\right) 
	+
	\frac{1}{2}
	\frac{\partial }{\partial S(\mathbf{p})^i}
	\left(
	a_\mathbf{p}
	\delta a_\mathbf{p}^\dag\,
	-
	\delta a_\mathbf{p}
	a_\mathbf{p}^\dag
	\right)
\end{align*}
and the analogous one for $b, b^\dag$. The second term, after integration by parts, gives the contribution to the integral 
\begin{align}
    - \frac{1}{2}
	\frac{\partial [S(\omega_\mathbf{p}) ]}{\partial S(\mathbf{p})^i}
	\left(
	a_\mathbf{p}
	\delta a_\mathbf{p}^\dag\,
	-
	\delta a_\mathbf{p}
	a_\mathbf{p}^\dag
	\right)
	+
	\frac{1}{2}
	\frac{\partial [\omega_\mathbf{p} ]}{\partial \mathbf{p}^i}
	\left(
	b_{\mathbf{p}}
	\delta b_{\mathbf{p}}^\dag
	-
	\delta b_{\mathbf{p}} b_{\mathbf{p}}^\dag
	\right)
\end{align}
and once again it eliminates the relative term in the computation of the boost charge.

For rotations, we can proceed in the same manner as for boost, and from
\begin{align}\label{dlorentz}
	\delta^R a_\mathbf{p}
	&=
	i \epsilon^{ijk} \rho_i
	\mathbf{p}_{j} \frac{\partial}{\partial \mathbf{p}^{k}}
	a_\mathbf{p}, 
\qquad
	\delta^R a_\mathbf{p}^\dag
	=
	i \epsilon^{ijk} \rho_i
	S(\mathbf{p})_{j} \frac{\partial}{\partial S(\mathbf{p})^{k}}
	a_\mathbf{p}^\dag, \nonumber \\
	\delta^R b_\mathbf{p}
	&=
	i \epsilon^{ijk} \rho_i
	\mathbf{p}_{j} \frac{\partial}{\partial \mathbf{p}^{k}}
	b_\mathbf{p}, 
\qquad
	\delta^R b_\mathbf{p}^\dag
	=
	i \epsilon^{ijk} \rho_i
	S(\mathbf{p})_{j} \frac{\partial}{\partial S(\mathbf{p})^{k}}
b_\mathbf{p}^\dag
\end{align}
we find the charge to have the form
\begin{align}
{\cal	M}_{i}=
		-\epsilon_i{}^{jk}
		\frac{1}{8}
		\int  d^3q \,
		\left(
		S(\mathbf{q})_{j}
		\frac{\partial a_{\mathbf{q}}^\dag}{\partial S(\mathbf{q})^{k}}
		a_{\mathbf{q}}
		-
		a_{\mathbf{q}}^\dag
		S(\mathbf{q})_{j}
		\frac{\partial a_{\mathbf{q}}}{\partial S(\mathbf{q})^{k}}
		+
		b_{\mathbf{q}}
		\mathbf{q}_{j}
		\frac{\partial b_{\mathbf{q}}^\dag}{\partial \mathbf{q}^{k}}
		-
		\mathbf{q}_{j}
		\frac{\partial b_{\mathbf{q}}}{\partial \mathbf{q}^{k}}
		b_{\mathbf{q}}^\dag
		\right).
\end{align}
All the deformed charges reduce to the undeformed ones in the limit $\kappa\rightarrow\infty$.

Once again, we can compute the field transformations related to the above creation/annihilation operators. Proceeding as in the undeformed case, assuming a dependence of $\zeta(p)$ only on $\mathbf{p}^2$, we get that the rotation tranformation of the spacetime field is the same as in the undeformed case. 

The boost transformation is however modified to
\begin{align}
    \delta^B\phi(x)  
    &=
    i \lambda_i\, x^i\frac{\partial }{\partial t}\, \phi(x) \nonumber \\
    &-
    i\lambda_i\, 
\int \frac{d^3p}{\sqrt{2\omega_\mathbf{p}}}\, \,
\,\zeta(p)\,
\Bigg\{
\omega_\mathbf{p}
\left[
\frac{1}{2}
\frac{1}{\omega_\mathbf{p}}
	\frac{\partial [\omega_\mathbf{p} ]}{\partial \mathbf{p}^i}
-
\frac{\sqrt{\omega_\mathbf{p}}}{\zeta(p) \omega_p}
\frac{\partial }{\partial \mathbf{p}^i}
\left(
\frac{\omega_\mathbf{p}}{\sqrt{ \omega_\mathbf{p}}}
\zeta(p)
\right)
\right]a_{\mathbf{p}}
e^{-i(\omega_\mathbf{p}t-\mathbf{p}\mathbf{x})}  \nonumber \\
&+
S(\omega_\mathbf{p})
\left[
\frac{1}{2}
\frac{1}{S(\omega_\mathbf{p})}
    \frac{\partial [S(\omega_\mathbf{p}) ]}{\partial S(\mathbf{p})^i}
-
\frac{\sqrt{\omega_\mathbf{p}}}{\zeta(p)S(\omega_p)}
\frac{\partial }{\partial S(\mathbf{p})^i}
\left(
\frac{S(\omega_\mathbf{p})}{\sqrt{ \omega_\mathbf{p}}}
\zeta(p)
\right)
\right]b_{\mathbf{p}}^\dag
e^{-i(S(\omega_\mathbf{p})t-S(\mathbf{p})\mathbf{x})}
\Bigg\},
\end{align}
\begin{align}
    \delta^B\phi^\dag(x)
    &=
    -i \lambda_i\, x^i\frac{\partial }{\partial t}\, \phi^\dag(x) \nonumber \\
    &+
    i\lambda_i\, 
\int \frac{d^3p}{\sqrt{2\omega_\mathbf{p}}}\, \,
\,\zeta(p)\,
\Bigg\{
\omega_\mathbf{p}
\left[
\frac{1}{2}
\frac{1}{\omega_\mathbf{p}}
	\frac{\partial [\omega_\mathbf{p} ]}{\partial \mathbf{p}^i}
-
\frac{\sqrt{\omega_\mathbf{p}}}{\zeta(p) \omega_p}
\frac{\partial }{\partial \mathbf{p}^i}
\left(
\frac{\omega_\mathbf{p}}{\sqrt{ \omega_\mathbf{p}}}
\zeta(p)
\right)
\right]
b_{\mathbf{p}}
e^{-i(\omega_\mathbf{p}t-\mathbf{p}\mathbf{x})}
\nonumber \\
&+
S(\omega_\mathbf{p})
\left[
\frac{1}{2}
\frac{1}{S(\omega_\mathbf{p})}
    \frac{\partial [S(\omega_\mathbf{p}) ]}{\partial S(\mathbf{p})^i}
-
\frac{\sqrt{\omega_\mathbf{p}}}{\zeta(p)S(\omega_p)}
\frac{\partial }{\partial S(\mathbf{p})^i}
\left(
\frac{S(\omega_\mathbf{p})}{\sqrt{ \omega_\mathbf{p}}}
\zeta(p)
\right)
\right]
a_{\mathbf{p}}^\dag
e^{-i(S(\omega_\mathbf{p})t-S(\mathbf{p})\mathbf{x})} 
\Bigg\}.
\end{align}
This can be further simplified  (we only treat the case of $\phi$ since evidently $\phi^\dag$ can be treated in the same way)
\begin{align}\label{deltaBphi}
    \delta^B\phi(x)  
    &=
    i \lambda_i\, x^i\frac{\partial }{\partial t}\, \phi(x) \nonumber \\
    &+
    i\lambda_i\, 
\int \frac{d^3p}{\sqrt{2\omega_\mathbf{p}}}\, \,
\Bigg\{
\omega_\mathbf{p}
\frac{\partial \zeta(p)}{\partial \mathbf{p}^i}
a_{\mathbf{p}}
e^{-i(\omega_\mathbf{p}t-\mathbf{p}\mathbf{x})}  \nonumber \\
&+
\zeta(p)
\left[
\frac{S(\mathbf{p})_i}{2S(\omega_\mathbf{p})}
-
\frac{1}{2}
\frac{p_+}{\kappa}\frac{\mathbf{p}_i}{\omega_\mathbf{p}}
+
\frac{p_+}{\kappa}
\frac{\omega_\mathbf{p}}{\zeta(p)}
\frac{\partial \zeta(p)}{\partial \mathbf{p}^i}
\right]b_{\mathbf{p}}^\dag
e^{-i(S(\omega_\mathbf{p})t-S(\mathbf{p})\mathbf{x})}
\Bigg\}.
\end{align}
To understand these additional terms, let us use eq. \eqref{convention}, and let us expand the coefficients up to first order in $1/\kappa$. We have \eqref{convention}
$$
    \zeta(p)^{-2} = \frac{1}{2\omega_{\mathbf{p}}}\, \frac{p_4}{\kappa}\, \left[
	\frac{|p_+|^3}{\kappa^3}
	\omega_{\mathbf{p}} - S(\omega_{\mathbf{p}})
	\right],	
$$
so that, to the leading order in deformation,
\begin{align}\label{conventionexpanded}
	\zeta(p) 
	\approx 
	1
	-
	\frac{1}{\kappa}
	\frac{2\omega_\mathbf{p}^2 + m^2}{4\omega_\mathbf{p}},
\end{align}
from which we have
\begin{align}
    \omega_\mathbf{p} \frac{\partial \zeta(p)}{\partial \mathbf{p}^i}
	\approx
	\omega_\mathbf{p}
	\frac{1}{\kappa}
	\left(
	-\frac{\mathbf{p}_i}{2\omega_\mathbf{p}}
	+
	m^2
	\frac{1}{4\omega_\mathbf{p}^2}
	\frac{\mathbf{p}_i}{\omega_\mathbf{p}}
	\right)
	=
	\frac{\mathbf{p}_i}{\kappa}
	\left(
	\frac{m^2}{4\omega_\mathbf{p}^2}
	-
	\frac{1}{2}
	\right)
\end{align}
and
\begin{align}
    \frac{S(\mathbf{p})_i}{2S(\omega_\mathbf{p})}
    -
    \frac{1}{2}
    \frac{p_+}{\kappa}\frac{\mathbf{p}_i}{\omega_\mathbf{p}}
    +
    \frac{p_+}{\kappa}
    \frac{\omega_\mathbf{p}}{\zeta(p)}
    \frac{\partial \zeta(p)}{\partial \mathbf{p}^i}
%   &\approx \nonumber \\
%    \frac{-\mathbf{p}_i\left(1 - \frac{\omega_\mathbf{p}}{\kappa}\right)}{-2\omega_\mathbf{p} + 2\frac{\mathbf{p}^2}{\kappa}}
%    -
%    \frac{\mathbf{p}_i}{2\omega_\mathbf{p}}
%    \left(1 + \frac{\omega_\mathbf{p}}{\kappa}\right)
%    &+
%    \left(1 + \frac{\omega_\mathbf{p}}{\kappa}\right)
%    \sqrt{2}
%    \left(
%    1 + \frac{3\omega_\mathbf{p}}{4\kappa}
%    \right)
%    \left(
%    -
%    \frac{3}{2\sqrt{2}}
%    \frac{\mathbf{p}_i}{\kappa}
%   \right) \nonumber \\
    &\approx
	\frac{\mathbf{p}_i}{\kappa}
	\left(
	-\frac{m^2 }{4\omega_\mathbf{p}^2}
	-
	1
	\right).
\end{align}
The boost transformations therefore become 
\begin{align}\label{LOphibo}
	\delta^B\phi(x)  
	&=
	i \lambda_i\, x^i\frac{\partial }{\partial t}\, \phi(x) 
	-
	i\lambda_i\, 
	\int \frac{d^3p}{\sqrt{2\omega_\mathbf{p}}}\, \,
	\Bigg\{
	\frac{\mathbf{p}_i}{\kappa}
	\left(
	\frac{m^2}{4\omega_\mathbf{p}^2}
	-
	\frac{1}{2}
	\right)
	a_{\mathbf{p}}
	e^{-i(\omega_\mathbf{p}t-\mathbf{p}\mathbf{x})}  \nonumber \\
	&+
	\frac{\mathbf{p}_i}{\kappa}
	\left(
	-\frac{m^2 }{4\omega_\mathbf{p}^2}
	-
	1
	\right)
	b_{\mathbf{p}}^\dag
	e^{-i(S(\omega_\mathbf{p})t-S(\mathbf{p})\mathbf{x})}
	\Bigg\},
\end{align}
\begin{align}\label{LOphidagbo}
	\delta^B\phi^\dag(x)
	&=
	-i \lambda_i\, x^i\frac{\partial }{\partial t}\, \phi(x)^\dag 
	+
	i\lambda^i\, 
	\int \frac{d^3p}{\sqrt{2\omega_\mathbf{p}}}\, \,
	\Bigg\{
	\frac{\mathbf{p}_i}{\kappa}
	\left(
	\frac{m^2}{4\omega_\mathbf{p}^2}
	-
	\frac{1}{2}
	\right)
	b_{\mathbf{p}}
	e^{-i(\omega_\mathbf{p}t-\mathbf{p}\mathbf{x})}
	\nonumber \\
	&+
	\frac{\mathbf{p}_i}{\kappa}
	\left(
	-\frac{m^2 }{4\omega_\mathbf{p}^2}
	-
	1
	\right)
	a_{\mathbf{p}}^\dag
	e^{-i(S(\omega_\mathbf{p})t-S(\mathbf{p})\mathbf{x})} 
	\Bigg\}.
\end{align}
Notice that, using Eq. \eqref{convention}, these additional translations are the same for $a^\dag$ and $b^\dag$.
We see that to the leading order the additional terms in Eq. \eqref{deltaBphi} correspond to energy-dependent translations, different for the positive and negative energy modes. In particular, for very energetic particle, for which $\omega_\mathbf{p}^2\gg m^2$, we have  to do with the infinitesimal translation with the parameter $\lambda^i/2\kappa$ for positive energy modes, and with the parameter  $\lambda^i/\kappa$ for the negative energy ones. 

These additional translations can be absorbed into redefinition of the boosts so as to make the spacetime boost transformation \eqref{deltaBphi} containing only the classical first term. But such change of the form of the boosts will change the form of the commutator of boost generators (cf. Eq. \eqref{comBB} further in the text), which we want to have the classical form. Therefore in what follows we will stick to the definition of boosts \eqref{dboost}.

This concludes the computation of the charges. Let us now  prove that they form a representation of the Poincar\'e algebra.

\subsection{The algebra of charges}

In the previous subsection we found ten Noether-conserved charges associated with symmetries of our deformed theory. Now we must check if the charges form a representation of the Poincar\'e algebra, i.e., if they satisfy the commutation relations:
\begin{align}
	[M_j, P_k] = i \epsilon_{jkl} P_l,
	\qquad 
	[M_i, P_0] = 0,
	\qquad 
	[N_j, P_k] = - i \eta_{jk} P_0,
	\qquad
	[N_j, P_0] = - i P_j,
\end{align}
\begin{align}
	[M_j, M_k] = i \epsilon_{jkl} M_l,
	\qquad
	[M_j, N_k] = i \epsilon_{jkl} N_l,
	\qquad
	[N_j, N_k] = -i \epsilon_{jkl} M_l.
\end{align}
 To compute the brackets we use the  creation/annihilation operators algebra \eqref{aadagalg}, and since from now on we will use exclusively the (formal) operator algebra, we do not use caret symbol\, $\hat{}$ to distinguish them.

To start with, we consider the boost and momenta commutator.
We have
\begin{align}\label{ansatz1comp}
	[\mathcal{N}_j, \mathcal{P}_0]
	&=-\, 
	\frac{i}{2}
	\int
	d^3p \, d^3q \,
	 S(\omega_p) S(\omega_q)
	\Bigg\{
	\left[
	\frac{\partial a_{\mathbf{p}}^\dag}{\partial S(\mathbf{p})^j} a_{\mathbf{p}},
	a_{\mathbf{q}}^\dag a_{\mathbf{q}}
	\right]
	-
	\left[
	a_{\mathbf{p}}^\dag
	\frac{\partial a_{\mathbf{p}}}{\partial S(\mathbf{p})^j} ,
	a_{\mathbf{q}}^\dag a_{\mathbf{q}} 
	\right]
	\Bigg\} \nonumber \\
	&-\omega_p \omega_q
	\Bigg\{
	\left[
	b_{\mathbf{p}}
	\frac{\partial b_{\mathbf{p}}^\dag}{\partial \mathbf{p}^j}  ,
	b_{\mathbf{q}}b_{\mathbf{q}}^\dag 
	\right]
	-
	\left[
	\frac{\partial  b_{\mathbf{p}}}{\partial \mathbf{p}^j}
	b_{\mathbf{p}}^\dag
	,
	b_{\mathbf{q}}b_{\mathbf{q}}^\dag 
	\right]
	\Bigg\}.
\end{align}
We can treat the $a$ and $b$ parts separately. Starting from the former, we first notice that as the result of the fact that $\left[
	a_{\mathbf{p}}^\dag
 a_{\mathbf{p}},
	a_{\mathbf{q}}^\dag a_{\mathbf{q}} 
	\right]=0$, the two commutators inside the curly bracket are equal and therefore it suffices to compute the first one. We find
\begin{align*}
    -\, 
i
	&\int
	d^3p \, d^3q \,
	 S(\omega_p) S(\omega_q)
	\left[
	\frac{\partial a_{\mathbf{p}}^\dag}{\partial S(\mathbf{p})^j} a_{\mathbf{p}},
	a_{\mathbf{q}}^\dag a_{\mathbf{q}}
	\right] \\= -\, 
	i
&	\int
	d^3p \, d^3q \,
	 S(\omega_p) S(\omega_q)\,\left(\frac{\partial a_{\mathbf{p}}^\dag}{\partial S(\mathbf{p})^j}a_{\mathbf{q}}\,\delta(\mathbf{p}-\mathbf{q} )-a_{\mathbf{q}}^\dag a_{\mathbf{p}}\,\frac{\partial }{\partial S(\mathbf{p})^j} \,\delta(\mathbf{p}-\mathbf{q} )\right)\\
	 =	i
&	\int
	d^3p \, S(p)_j\, a_{\mathbf{p}}^\dag a_{\mathbf{p}},
\end{align*}	
where in the last step we integrated by parts a couple of times. The $b$	commutator can be handled in exactly the same way, which gives us the desired result
\begin{equation}\label{NPialg}
 	[\mathcal{N}_j, \mathcal{P}_0] =-i    \mathcal{P}_j.
\end{equation}	
In exactly the same way one shows that	
\begin{equation}\label{NP0alg}
 	[\mathcal{N}_j, \mathcal{P}_k] =-i \delta_{jk}   \mathcal{P}_0.
\end{equation}	
  
\noindent The commutators of rotation and translation charges can be computed analogously.

Now let us turn to the Lorentz sector. Let us consider, as an example, the commutator
\begin{align}
	[\mathcal{N}_j, \mathcal{N}_k]
	&=
	\frac{i}{4}
	\int d^3p \, d^3q
	\Bigg[
	 S(\omega_p)
	\Bigg\{
	\frac{\partial a_{\mathbf{p}}^\dag}{\partial S(\mathbf{p})^j}
	a_{\mathbf{p}}
	-
	a_{\mathbf{p}}^\dag
	\frac{\partial a_{\mathbf{p}}}{\partial S(\mathbf{p})^j}
	\Bigg\}
	+
	\omega_p
	\Bigg\{
	b_{\mathbf{p}} 
	\frac{\partial b_{\mathbf{p}}^\dag}{\partial \mathbf{p}^j}
	-
	\frac{\partial b_{\mathbf{p}}}{\partial \mathbf{p}^j} b_{\mathbf{p}}^\dag
	\Bigg\}, \nonumber\\
	&
	S(\omega_q)
	\Bigg\{
	\frac{\partial a_{\mathbf{q}}^\dag}{\partial S(\mathbf{q})^k}
	a_{\mathbf{q}}
	-
	a_{\mathbf{q}}^\dag
	\frac{\partial a_{\mathbf{q}}}{\partial S(\mathbf{q})^k}
	\Bigg\}
	+
	 \omega_q
	\Bigg\{
	b_{\mathbf{q}} 
	\frac{\partial b_{\mathbf{q}}^\dag}{\partial \mathbf{q}^k}
	-
	\frac{\partial b_{\mathbf{q}}}{\partial \mathbf{q}^k} b_{\mathbf{q}}^\dag
	\Bigg\}
	\Bigg].
\end{align}
The procedure is basically the same as before. 
Once again, the mixed commutators will go away and we are only left with the commutators between the $a$'s and the $b$'s. Expanding the terms containing $a, a^\dag$ we notice that two terms cancel each other, and the remaining two are equal except for the exchanges $\mathbf{p} \leftrightarrow \mathbf{q}$ and $j\leftrightarrow k$, but we can just rename the integration variables in such a way that the second one and the first are the same except for $j \leftrightarrow k$. We can then proceed as before integrating by parts and applying the Dirac delta when appropriate. We get  the final result 
\begin{align}
    [\mathcal{N}_i, \mathcal{N}_j]=
		\frac{1}{4}
		\int  d^3q \,
		\left(
		S(\mathbf{q})_{[i}
		\frac{\partial a_{\mathbf{q}}^\dag}{\partial S(\mathbf{q})^{j]}}
		a_{\mathbf{q}} 
		-
		a_{\mathbf{q}}^\dag
		S(\mathbf{q})_{[i}
		\frac{\partial a_{\mathbf{q}}}{\partial S(\mathbf{q})^{j]}}
		+
		b_{\mathbf{q}}
		\mathbf{q}_{[i}
		\frac{\partial b_{\mathbf{q}}^\dag}{\partial \mathbf{q}^{j]}}
		-
		\mathbf{q}_{[i}
		\frac{\partial b_{\mathbf{q}}}{\partial \mathbf{q}^{j]}}
		b_{\mathbf{q}}^\dag
		\right)
\end{align}
which we can rewrite in the desired final form
\begin{align}\label{comBB}
	[\mathcal{N}_j, \mathcal{N}_k]
	=
	-i
	\tensor{\epsilon}{_{jk}^l}
	\mathcal{M}_l.
\end{align}
Other Lorentz charges commutators can be computed in an analogous way and this completes our check that the conserved Noether charges indeed form a representation of the Poincar\'e algebra.

Let us summarize the findings so far. We found that our theory possesses ten conserved Noether charges that under commutator form the Poincar\'e algebra. By Noether theorem this tells us  that the theory is Poincar\'e-invariant. As we will see in Sect.\ \ref{SectV} this is weak form of Poincar\'e invariance is not sufficiently powerful to deduce the CPT theorem from it. In fact, we will see in the net section that in spite of the Poincar\'e invariance the theory breaks the CPT symmetry in a subtle way.

\section{Charge symmetry operator and its properties}\label{SecIV}

As discussed in our previous paper \cite{Arzano:2020jro},  the charge conjugation operator  $C$ is given by
\begin{align}
	\mathcal{C} =
	\int \, d^3p
	\,
	(
	b^\dag_{\mathbf{p}}
	a_\mathbf{p}
	+
	a^\dag_{\mathbf{p}}
	b_{\mathbf{p}}
	).
\end{align}
It is easy to see that this operator satisfies the canonical $C$ property
\begin{align}
	\mathcal{C} a^\dag |0\rangle = b^\dag |0\rangle
	\qquad
	\mathcal{C} b^\dag |0\rangle = a^\dag |0\rangle,
\end{align}
i.e., it exchanges particles with antiparticles.
Moreover, $C=C^\dag$ and $C^2 = 1$. 

In the undeformed case it is easy to see that $[\mathcal{N}_i, \mathcal{C}]_{\text{undef}}=0$. In fact, using the canonical boost generator
\begin{align}
    \mathcal{N}_i =
	\frac{1}{2}
	\int
	d^3p \,
	i\,\omega_p
	\Bigg\{
	\frac{\partial a_{\mathbf{p}}^\dag}{\partial \mathbf{p}^i}
	a_{\mathbf{p}}
	-
	a_{\mathbf{p}}^\dag
	\frac{\partial a_{\mathbf{p}}}{\partial \mathbf{p}^i}
	\Bigg\}
	+
	i \omega_p
	\Bigg\{
	b_{\mathbf{p}}
	\frac{\partial b_{\mathbf{p}}^\dag}{\partial \mathbf{p}^i}
	-
	\frac{\partial b_{\mathbf{p}}}{\partial \mathbf{p}^i} b_{\mathbf{p}}^\dag
	\Bigg\} 
	=
	\mathcal{N}_i^a
	+
	\mathcal{N}_i^b
\end{align}
and the commutation relation between $a, a^\dag$ and $b, b^\dag$ one gets 
\begin{align}
	[\mathcal{N}_j, \mathcal{C}]_{\text{undef}}
	&=
	\frac{i}{2}
	\int \, d^3p \,
	\Bigg\{
	\omega_p
	\left[
	\frac{\partial a_{\mathbf{p}}}{\partial \mathbf{p}^j}
	b^\dag_{\mathbf{p}}
	-
	a_{\mathbf{p}}
	\frac{\partial b^\dag_{\mathbf{p}} }{\partial \mathbf{p}^j}
	+
	\frac{\partial a_{\mathbf{p}}^\dag}{\partial \mathbf{p}^j}
	b_{\mathbf{p}}
	-
	a_{\mathbf{p}}^\dag
	\frac{\partial b _{\mathbf{p}}}{\partial \mathbf{p}^j}
	\right] \nonumber \\
	&+
	\omega_p
	\left[
	\frac{\partial b_{\mathbf{p}}^\dag}{\partial \mathbf{p}^j}
	a_\mathbf{p}
	-
	b_{\mathbf{p}}^\dag
	\frac{\partial a_\mathbf{p}}{\partial \mathbf{p}^j}
	+
	\frac{\partial b_{\mathbf{p}}}{\partial \mathbf{p}^j}
	a^\dag_{\mathbf{p}}
	-
	b_{\mathbf{p}}
	\frac{\partial a_\mathbf{p}^\dag}{\partial \mathbf{p}^j}
	\right]
	\Bigg\}  
	=0,
\end{align}
which is due to the fact that the terms in the two square brackets (which correspond to the integrands of $[\mathcal{N}_j^a, \mathcal{C}]$ and $[\mathcal{N}_j^b,\mathcal{C}]$ respectively) are equal and opposite.

In the deformed case the boost is given by eq. \eqref{defboostfromsymp} and its  commutator with charge congugation will be given by
\begin{align}
	[\mathcal{N}_j, \mathcal{C}]
	&=
	\frac{i}{2}
	\int \, d^3p \,
	\Bigg\{
	S(\omega_p)
	\left[
	\frac{\partial a_{\mathbf{p}}}{\partial S(\mathbf{p})^j}
	b^\dag_{\mathbf{p}}
	-
	a_{\mathbf{p}}
	\frac{\partial b^\dag_{\mathbf{p}} }{\partial S(\mathbf{p})^j}
	+
	\frac{\partial a_{\mathbf{p}}^\dag}{\partial S(\mathbf{p})^j}
	b_{\mathbf{p}}
	-
	a_{\mathbf{p}}^\dag
	\frac{\partial b _{\mathbf{p}}}{\partial S(\mathbf{p})^j}
	\right] \nonumber \\
	&+
	\omega_p
	\left[
	\frac{\partial b_{\mathbf{p}}^\dag}{\partial \mathbf{p}^j}
	a_\mathbf{p}
	-
	b_{\mathbf{p}}^\dag
	\frac{\partial a_\mathbf{p}}{\partial \mathbf{p}^j}
	+
	\frac{\partial b_{\mathbf{p}}}{\partial \mathbf{p}^j}
	a^\dag_{\mathbf{p}}
	-
	b_{\mathbf{p}}
	\frac{\partial a_\mathbf{p}^\dag}{\partial \mathbf{p}^j}
	\right]
	\Bigg\}
\end{align}
which is  not-zero because the two square brackets are not equal and opposite-sign anymore\footnote{Notice that one can also easily verify that $[\mathcal{P}_\mu, \mathcal{C}]\neq 0$, which could also have experimental consequences.}. This is the major new prediction of the $\kappa$-deformed field theory that, as we will discus in details below in Sect.\ \ref{Sect6}, can be, in principle, observable. Let us discuss the origin and consequences of this in some more details.

As seen above (cf. also Ref.\ \cite{Arzano:2020jro}) in the case of $\kappa$-deformed field theory  boosts act in an unexpected way. Let us consider momenta carried by one-particle states. They can be computed acting on the states with the momentum charge operators
\begin{align}
	\mathcal{P}_i |\mathbf{p}\rangle_a
	=
	-S(\mathbf{p})_i |\mathbf{p}\rangle_a
	\qquad
	\mathcal{P}_i |\mathbf{p}\rangle_b
	=
	\mathbf{p}_i |\mathbf{p}\rangle_b\nonumber\\
	\mathcal{P}_0 |\mathbf{p}\rangle_a
	=
	-S(\omega_p) |\mathbf{p}\rangle_a
	\qquad
	\mathcal{P}_0 |\mathbf{p}\rangle_b
	=
	\omega_p |\mathbf{p}\rangle_b
\end{align}
where $|\mathbf{p}\rangle_a$ is the one $a$-particle state labelled by $\mathbf{p}$. Moreover,
\begin{align}
	\mathcal{C} |\mathbf{p}\rangle_b
	=
	\mathcal{C}b^\dag_{\mathbf{p}*}\mathcal{C}^{-1}\mathcal{C}|0\rangle
	=
	a^\dag_{\mathbf{p}}|0\rangle
	=
	|\mathbf{p}\rangle_a.
\end{align}
Than, in particular, for a particle at rest
\begin{align}
	\mathcal{P}_i |\mathbf{p}=0\rangle_a
	=
	0
	\qquad
	\mathcal{P}_i |\mathbf{p}=0\rangle_b
	=
	0\nonumber\\
	\mathcal{P}_0 |\mathbf{p}\rangle_a
	=
	M |\mathbf{p}\rangle_a
	\qquad
	\mathcal{P}_0 |\mathbf{p}\rangle_b
	=
	M |\mathbf{p}\rangle_b
\end{align}
from which it follows that the particle and antiparticle have the same rest mass.

This changes when we apply the boost. Making use of the algebra of boost and translation generators \eqref{NPialg}, \eqref{NP0alg} for an infinitesimal boost parameter $\lambda^j$ we find
\begin{align*}
-i\lambda^j	\mathcal{N}_j\,	\mathcal{P}_i |\mathbf{p}\rangle_a &=iS(p)_i \lambda^j	\mathcal{N}_j	 |\mathbf{p}\rangle_a =-i\mathcal{P}_i \,\lambda^j	\mathcal{N}_j	 |\mathbf{p}\rangle_a + \lambda_{i} S(\omega_\mathbf{p})|\mathbf{p}\rangle_a\\
-i\lambda^j	\mathcal{N}_j\,	\mathcal{P}_0 |\mathbf{p}\rangle_a &=iS(\omega_\mathbf{p}) \lambda^j	\mathcal{N}_j	 |\mathbf{p}\rangle_a =-i\mathcal{P}_0 \,\lambda^j	\mathcal{N}_j	 |\mathbf{p}\rangle_a + \lambda^{i} S(p)_i |\mathbf{p}\rangle_a
\end{align*}
which can be solved by 
\begin{equation}\label{Naction}
  -i  \lambda^j	\mathcal{N}_j	 |\mathbf{p}\rangle_a = |\mathbf{p}+ \bm{\lambda}\omega_\mathbf{p} \rangle_a. 
\end{equation}
One easily checks that the  boost acts on $|\mathbf{p}\rangle_b$ in exactly the same way. Having the infinitesimal boost action \eqref{Naction} we can  find the action of a finite boost, in direction $1$ say, characterized by the rapidity $\xi$ (related to the Lorentz boost $\gamma=\cosh\xi$)
\begin{align}\label{finiteboost}
  	 |p_1,p_2,p_3\rangle_a &\mapsto |\cosh\xi\, p_1 +\sinh\xi\, \omega_\mathbf{p},p_2,p_3\rangle_a\nonumber\\
  	 |p_1,p_2,p_3\rangle_b &\mapsto |\cosh\xi\, p_1 +\sinh\xi\, \omega_\mathbf{p},p_2,p_3\rangle_b.
\end{align}
In particular, when we boost a particle, initially at rest, we get
\begin{align}\label{finiteboost0}
  	 |\mathbf{p}=0\rangle_a &\mapsto |M\sinh\xi,0,0\rangle_a\nonumber\\
  	 |\mathbf{p}=0\rangle_b &\mapsto |M\sinh\xi,0,0\rangle_b
\end{align}
and such boosted states will carry momenta and energies
\begin{align}
	\mathcal{P}_1 |M\sinh\xi,0,0\rangle_a
&	=
	-S(M\sinh\xi)_1|M\sinh\xi,0,0\rangle_a
\nonumber\\
	\mathcal{P}_0 |M\sinh\xi,0,0\rangle_a
&	=
-S(M\cosh\xi) |M\sinh\xi,0,0\rangle_a
\end{align}
and 
\begin{align}
	\mathcal{P}_1 |M\sinh\xi,0,0\rangle_b
&	=
	M\sinh\xi|M\sinh\xi,0,0\rangle_b
\nonumber\\
	\mathcal{P}_0 |M\sinh\xi,0,0\rangle_b
&	=
M\cosh\xi |M\sinh\xi,0,0\rangle_b 
\end{align}
with
\begin{align}\label{eqn:SofP}
    -S(M\sinh\xi)_1 &= \frac{\kappa M\sinh\xi}{M\cosh\xi  +\sqrt{\kappa^2 +M^2}}\,,\nonumber\\
    -S(M\cosh\xi)&=M \cosh\xi - \frac{M^2\sinh^2\xi}{M\cosh\xi  +\sqrt{\kappa^2 +M^2}}
\end{align}
To conclude, we see that although particles and antiparticles have equal rest masses, since the charge conjugation operator does not commute with the boost, acting on the particle state of a given, nonzero momentum charge conjugation produces an antiparticle state of a different momentum. We will discuss the phenomenological implications of this effect in Sect.\ VI.

\section{On relations between CPT and Poincar\'e invariance}\label{SectV}

The above results and previous work \cite{Arzano:2020jro} provide us with an explicit model of a Poincar\'e-invariant theory that breaks CPT in a subtle way. This stands in an apparent contradiction with Jost-Wightman-Greenberg theorem \cite{Jost:1957zz}, \cite{Wightman:1963deu}, \cite{Greenberg:2002uu} stating that  CPT invariance violation results in Lorentz invariance breaking and vice versa (see eg., \cite{Greenberg:2003nv}, \cite{Lehnert:2016zym} for  reviews and \cite{Duetsch:2012sd} and 
	\cite{Chaichian:2011fc} for counterarguments). In this section we recall briefly the reasoning leading to the Jost-Wightman-Greenberg theorem and then argue that it is not applicable in the $\kappa$-deformed context.

The central quantities of interest are the Wightman functions
\begin{align}\label{Wightmanf}
	W(x_1, \dots, x_n) = \langle 0 | \phi_1(x_1) \dots \phi_n(x_n) |0\rangle
\end{align}
and their properties under Poincar\'e transformations. The CPT theorem can be expressed in terms of the Wightman function in the form of the condition
	\begin{align}\label{CPTTheorem}
		\langle 0 | \phi_1(x_1) \dots \phi_n(x_n) |0\rangle
		=
		(-1)^L 
		(i)^F
		\langle 0 | \phi_1^\dag(-x_1) \dots \phi_n^\dag(-x_n) |0\rangle^*
	\end{align}
	and from this one can read off the CPT transformation 
	\begin{align}\label{CPTCond}
		\theta \phi(x) \theta^\dag
		=
		(-1)^l i^f \phi^\dag(-x). 
	\end{align}
The powers $L$, $F$ are defined below, while $l$ and $f$ refer to the number of dotted and undotted indices in the field representation used in \cite{Greenberg:2003nv}.

In the following, we sketch the line of reasoning that underlies both Greenberg's argument and Jost theorem. %The proof of the Jost theorem, showing that \eqref{CPTTheorem} follows from Poincar\'e symmetry proceeds in the following steps \cite{Greenberg:2003nv}:
\begin{itemize}
\item[1)] The argument presented in \cite{Greenberg:2002uu} relating CPT invariance to Lorentz symmetry starts from the proof that, because of out-of-cone Lorentz invariance of the theory, one has 
\begin{align}\label{Jostcond}
    W(x_1, \dots, x_n) = W(x_n, \dots, x_1).
\end{align}
This condition is called \textit{weak local commutativity}. Then one can use Jost theorem, which states that equation \eqref{Jostcond}\footnote{More precisely,  \eqref{Jostcond} must be valid in the neighbour of the Jost points, which are described in the later points.} is equivalent to CPT invariance of the theory, i.e. to  eq.\ \eqref{CPTTheorem}. In the following points, we sketch the proof of Jost theorem presented in more details in \cite{Jost:1957zz}, \cite{Wightman:1963deu}, \cite{Greenberg:2002uu} \cite{Greenberg:2003nv}, \cite{Lehnert:2016zym}.

	\item[2)] The key idea in Jost theorem is to use the group $SL(2, \mathbb{C})$, which is the double covering of the proper orthochronous Lorentz group $L_+^{\uparrow}$. One uses $SL(2, \mathbb{C})$ instead of $L_+^{\uparrow}$ because spacetime inversion $PT = -1$ is connected to the identity in $SL(2, \mathbb{C})$, while it is not connectd in $L_+^{\uparrow}$. 
	
	\item[3)] To use the complex Lorentz group, one needs to analytically continue the Wightman functions, i.e., the vacuum expectation values of products of fields. As a consequence of translational invariance, instead of the canonical $W^{(n)}(x_1, \dots, x_n)$, one employs  equivalently the function $\tilde{W}(x_1-x_2, \dots, x_{n-1}-x_n) := \tilde{W}(\xi_1, \dots, \xi_{n-1})$ where $\xi_j = x_j - x_{j+1}$. In order to obtain an analytic continuation of $\tilde{W}$, we promote each difference $\xi_j$ to a complex variable, and in particular
	\begin{align}
		z_j= \xi_j -i\eta_j. 
	\end{align}
	The introduction of the $-i\eta_j$ transforms the distribution $\tilde{W}(\xi_1, \dots, \xi_{n-1})$ into an analytic function $\tilde{W}(z_1, \dots, z_{n-1})$.
	
	Notice that the use of the differences $\xi_j$ is a \textit{necessary step} for the proof. In fact, the whole argument rests on the existence of real points in the analyticity domain of the analytic function $\tilde{W}(z_1, \dots, z_{n-1})$. These real points $\{\xi_j\}$ are the so called Jost points, and are defined by the condition that the sum
	\begin{align}
		s:=\sum_j c_j \xi_j
	\end{align}
	is spacelike, where $c_j\geq 0$ and $\sum_j c_j >0$. Of course, since $s$ must be spacelike for any observers, it must be defined in terms of $\xi_j$ and not $x_j$. 
	
	\item[4)] The analytic functions  $\tilde{W}(z_1, \dots, z_{n-1})$ are  defined only in a domain which does \textit{not} include real points. However, because of Lorentz invariance with respect to the double covering group $SL(2,\mathbb{C})$, we have for any (complex) Lorentz transformation $\Lambda$
	\begin{align}\label{complexlorentz}
		\tilde{W}(\Lambda z_1, \dots, \Lambda z_{n-1})
		=
		\tilde{W}(z_1, \dots, z_{n-1})
	\end{align}
	This allows us to enlarge the domain, and the new domain does contain the Jost points. Furthermore, because of the fact that $SL(2,\mathbb{C})$ contains spacetime inversion ${\mathcal PT}$, there is some $\Lambda$ such that 
	\begin{align}
		\tilde{W}(\Lambda z_1, \dots, \Lambda z_{n-1})
		=
		(-1)^L\tilde{W}(-z_1, \dots, -z_{n-1})
	\end{align}
	for some $L$, see \cite{Greenberg:2003nv}. Writing the above relation \textit{at the Jost points}, using the vacuum expectation values instead of the Wightman functions and  \eqref{complexlorentz} we get\footnote{The fields $\phi_i$ in this expression can be any kind of field, boson or fermion.}
	\begin{align}\label{identity1}
		\langle 0 | \phi_1(x_1) \dots \phi_n(x_n) |0\rangle
		=
		(-1)^L 
		\langle 0 | \phi_1(-x_1) \dots \phi_n(-x_n) |0\rangle.
	\end{align}

	\item[5)]
	There is an  important subtlety in the above equality. In fact, the equality merely states that the value of the function on the LHS is the same as the one on the RHS.	However (essentially because of the antihermitian nature of the $\mathcal T$ operator) the function $(-1)^L 
	\langle 0 | \phi_1(-x_1) \dots \phi_n(-x_n) |0\rangle$
	has a different domain of definition than the function
	$\langle 0 |  \phi_1(x_1) \dots \phi_n(x_n) |0\rangle$. The way to solve this is to permute the fields inside the expectation value on the RHS of eq. \eqref{identity1}. This is where the hypothesis \eqref{Jostcond} comes in, because it allows us to perform this field re-ordering which is essential for the proof.
	
%	Notice that weak local commutativity is not needed for the theorem if, instead of considering just Wightman functions, one uses the time ordered products of Wightman functions. 

	\item[6)] With the weak local commutativity taken into account, eq.\ \eqref{identity1} becomes
	\begin{align}\label{jostpointsvev}
		\langle 0 | \phi_1(x_1) \dots \phi_n(x_n) |0\rangle
		=
		(-1)^L 
		(i)^F
		\langle 0 | \phi_n(-x_n) \dots \phi_1(-x_1) |0\rangle.
	\end{align}
	where $F$ accounts for the fermions in the expectation value. Now the functions on both sides have the same domain of definition, and since they are analytic and coincide in some open subset of their domain (i.e.\ in the neighbour(s) of  Jost points) the above equality implies the same equality of Wightman functions across the whole domain. Therefore we have
	\begin{align}
		\tilde{W}(z_1, \dots, z_n)
		=
		(-1)^L 
		(i)^F
		\tilde{W}(-z_n, \dots, -z_1). 
	\end{align}
	At this point, we can also take the limit $n_j \rightarrow 0$ and the analytic functions become once again distributions, and we have
	\begin{align}\label{finalwightman}
		\tilde{W}(\xi_1, \dots, \xi_n)
		=
		(-1)^L 
		(i)^F
		\tilde{W}(-\xi_n, \dots, -\xi_1). 
	\end{align}
	Notice that we could not take this limit before using the weak local commutativity because the domains of definitions of the functions were different. Therefore, we would have been stuck to the complex case, without the possibility of taking the limit and coming back to the real (physical) case. 

	\item[7)] We can write eq. \eqref{finalwightman} in terms of vacuum expectation value as
	\begin{align}
		\langle 0 | \phi_1(x_1) \dots \phi_n(x_n) |0\rangle
		=
		(-1)^L 
		(i)^F
		\langle 0 | \phi_n(-x_n) \dots \phi_1(-x_1) |0\rangle
	\end{align}
	Notice that this equation is formally equivalent to eq. \eqref{jostpointsvev}, but while eq. \eqref{jostpointsvev} only made sense in a neighbour of Jost points (i.e. the associated Wightman function was an analytic function) here everything is real (i.e. the Wightman functions are again distributions). One can restore the order of the fields through hermitian conjugation on the RHS obtaining
	\begin{align}
		\langle 0 | \phi_1(x_1) \dots \phi_n(x_n) |0\rangle
		=
		(-1)^L 
		(i)^F
		\langle 0 | \phi_1^\dag(-x_1) \dots \phi_n^\dag(-x_n) |0\rangle^*
	\end{align}
which is exactly the desired CPT theorem \eqref{CPTTheorem}.

	\item[8)] With the points $1)$ to $6)$ one has proved that (assuming Lorentz symmetry) the validity of the weak local commutativity at Jost points \eqref{Jostcond} implies CPT symmetry. To show the other direction one can just use the same steps in reverse order, and therefore weak local commutativity in the neighbour of Jost points is equivalent to CPT symmetry.
\end{itemize}

Let us now consider these steps in the context of $\kappa$-deformed theory. The starting point will be the construction of deformed Wightman function. It is natural to postulate that it should take the form analogous to \eqref{Wightmanf} with the ordinary products replaced with the star ones
\begin{align}\label{Wightmanfkappa}
	W_\kappa(x_1, \dots, x_n) = \langle 0 | \phi_1(x_1)\star \dots\star \phi_n(x_n) |0\rangle
\end{align}
Already as it stands this formula is problematic because the Fock space construction does not work in the deformed context (see \cite{ArzanoKowalski}). However the one particle spaces are well defined and we can consider the two-point functions, for which, using the field definitions \eqref{fa} and \eqref{fb} and the commutator algebra \eqref{aadagalg} one can check that
\begin{equation}\label{CPTTdef}
    \langle 0 | \phi^\dag(x_1) \star \phi(x_2) |0\rangle
		=
		\langle 0 | \phi(-x_1) \star \phi^\dag(-x_2) |0\rangle^*
\end{equation}
which shows that the two point functions are $CPT$- invariant. Notice the important fact that in verifying eq. \eqref{CPTTdef} we assumed the following commutator between two different coordinates
\begin{align}
    [x^0, \mathbf{y}^i] = \frac{i}{\kappa}\mathbf{y}^i.
\end{align}
Furthermore, one can also check that
\begin{equation}\label{Lorentz}
    \langle 0 | \phi^\dag(x_1) \star \phi(x_2) |0\rangle
		=
		\langle 0 | \phi(x_2) \star \phi^\dag(x_1) |0\rangle
\end{equation}
which shows out-of-cone Lorentz invariance for spacelike intervals $(x_2-x_1)$. However, because of the presence of the deformation, eq.\ \eqref{Lorentz} does not imply \eqref{CPTTdef}, nor vice versa. In fact, notice that in his paper \cite{Greenberg:2002uu} Greenberg uses the Lorentz invariance to show that eq. \eqref{Lorentz} holds, and then proceeds to use Jost theorem (for which the validity of eq. \eqref{Lorentz} is an essential point) to show that from \eqref{Lorentz} follows \eqref{CPTTdef}. In our case, however, one can see that the deformed counterpart of the crucial property of Wightman functions in eq.\ \eqref{complexlorentz} does not hold anymore. In fact, using eq. \eqref{conventionexpanded}, \eqref{LOphibo}, \eqref{LOphidagbo} one can show that to the leading order in $1/\kappa$
\begin{align}
	\langle 0 | \Lambda \phi^\dag (x_1) \star \Lambda \phi(x_2) |0\rangle
	&\approx
	\langle 0 | \phi^\dag (x_1) \star \phi(x_2) |0\rangle
	\nonumber \\
	&
	-i \lambda_j\, x^j_1\frac{\partial }{\partial x^0_1}\,
	\langle 0 | \phi^\dag (x_1) \star \phi(x_2) |0\rangle 
	+
	i \lambda_j\, x^j_2\frac{\partial }{\partial x^0_2}\,
	\langle 0 | \phi^\dag (x_1) \star \phi(x_2) |0\rangle \nonumber \\ 
	&+
	\int \, \frac{d^3p}{2\omega_\mathbf{p}} \,
	e^{i\mathbf{p}(\mathbf{x}_1 - \mathbf{x}_2)}
	e^{ip_0(x^0_1 - x^0_2)} 
	\left[
	i\lambda^j
	\frac{\mathbf{p}_j}{\kappa}
	\left(
	\frac{m^2}{2\omega_\mathbf{p}^2}
	+
	\frac{1}{2}
	\right)
	\right]
\end{align}
We first consider the terms
\begin{align}\label{remaining}
	-i \lambda_j\, x^j_1\frac{\partial }{\partial x^0_1}\,
	\langle 0 | \phi^\dag (x_1) \star \phi(x_2) |0\rangle 
	+
	i \lambda_j\, x^j_2\frac{\partial }{\partial x^0_2}\,
	\langle 0 | \phi^\dag (x_1) \star \phi(x_2) |0\rangle
\end{align}
Using eq. \eqref{CPTTdef} this becomes
\begin{align}
	-i \lambda_j\, x^j_1\frac{\partial }{\partial x^0_1}\,
	\langle 0 | \phi^\dag (x_1) \star \phi(x_2) |0\rangle 
	+
	i \lambda_j\, x^j_2\frac{\partial }{\partial x^0_2}\,
	\langle 0 | \phi(-x_1) \star \phi^\dag(-x_2) |0\rangle
\end{align}
and because of equations \eqref{fa},\eqref{fb} one can verify that
\begin{align}
	x^i_1\frac{\partial }{\partial x^0_1} \phi^\dag(x_1)
	=
	x^i_2\frac{\partial }{\partial x^0_2} \phi^\dag(-x_2)
\end{align}
and therefore
\begin{align}
	-i \lambda_j\, x^j_1\frac{\partial }{\partial x^0_1}\,
	\langle 0 | \phi^\dag (x_1) \star \phi(x_2) |0\rangle 
	+
	i \lambda_j\, x^j_2\frac{\partial }{\partial x^0_2}\,
	\langle 0 | \phi^\dag (x_1) \star \phi(x_2) |0\rangle
	=0
\end{align}
and the boost transformation property for the two point function becomes
\begin{align}\label{nnlstep}
	\langle 0 | \Lambda \phi^\dag (x_1) \star \Lambda \phi(x_2) |0\rangle
	&\approx
	\langle 0 | \phi^\dag (x_1) \star \phi(x_2) |0\rangle \nonumber \\
	&+
	\int \, \frac{d^3p}{2\omega_\mathbf{p}} \,
	e^{i\mathbf{p}(\mathbf{x}_1 - \mathbf{x}_2)}
	e^{i\omega_\mathbf{p}(x^0_1 - x^0_2)} 
	\left[
	i\lambda^j
	\frac{\mathbf{p}_j}{\kappa}
	\left(
	\frac{m^2}{2\omega_\mathbf{p}^2}
	+
	\frac{1}{2}
	\right)
	\right]
\end{align}
We now treat the second term on the RHS. First of all, recall that we are performing all the computations at equal times, so that we only need to consider 
\begin{align}
	\int \, \frac{d^3p}{2\omega_\mathbf{p}} \,
	e^{i\mathbf{p}(\mathbf{x}_1 - \mathbf{x}_2)}
	\left[
	i\lambda^j
	\frac{\mathbf{p}_j}{\kappa}
	\left(
	\frac{m^2}{2\omega_\mathbf{p}^2}
	+
	\frac{1}{2}
	\right)
	\right]
\end{align}
Calling $\Delta \mathbf{x} = \mathbf{x}_1 - \mathbf{x}_2$ we have
\begin{align}
	-2\pi 
	\frac{1}{i}
	\frac{\partial}{\partial \Delta \mathbf{x}^j} 
	\int \, \frac{d|\mathbf{p}| \mathbf{p}^2 d \cos\theta}{2\omega_\mathbf{p}} \,
	e^{i|\mathbf{p}||\Delta \mathbf{x}| \cos\theta}
	\left[
	i\lambda^j
	\frac{1}{\kappa}
	\left(
	\frac{m^2}{2\omega_\mathbf{p}^2}
	+
	\frac{1}{2}
	\right)
	\right]
\end{align}
Performing the angular integral one gets
\begin{align}
	\frac{\pi}{|\Delta \mathbf{x}|} 
	\frac{\lambda^j}{\kappa}
	\frac{1}{i}
	\frac{\partial}{\partial \Delta \mathbf{x}^j} 
	\int_{-\infty}^\infty \, \frac{d|\mathbf{p}| |\mathbf{p}|}{2\omega_\mathbf{p}} \,
	e^{i|\mathbf{p}||\Delta \mathbf{x}|}
	\left(
	\frac{m^2}{\omega_\mathbf{p}^2}
	+
	1
	\right)
\end{align}
One can easily check that, since $m>0, |\Delta \mathbf{x}|>0$, we have
\begin{align}
	\frac{m^2}{2}
	\int_{-\infty}^\infty \, d|\mathbf{p}|  \,
	e^{i|\mathbf{p}||\Delta \mathbf{x}|}
	\frac{|\mathbf{p}|}{(m^2 + \mathbf{p}^2)^{3/2}}
	=
	i m^2
	|\Delta \mathbf{x}| K_0(m|\Delta \mathbf{x}|)
\end{align}
\begin{align}\label{ft2}
	\frac{1}{2}
	\int_{-\infty}^\infty \, d|\mathbf{p}|  \,
	e^{i|\mathbf{p}||\Delta \mathbf{x}|}
	\frac{|\mathbf{p}|}{(m^2 + \mathbf{p}^2)^{1/2}}
	=
	i m
	K_1(m|\Delta \mathbf{x}|)
\end{align}
where $K_0(x), K_1(x)$ are the modified Bessel functions of the second kind. The additional term in eq. \eqref{nnlstep} therefore becomes
\begin{align}
	\frac{\pi}{|\Delta \mathbf{x}|} 
	\frac{\lambda^j}{\kappa}
	\frac{\partial m|\Delta \mathbf{x}|}{\partial \Delta \mathbf{x}^j} 
	\frac{\partial}{\partial m|\Delta \mathbf{x}|}
	\left(
	m^2
	|\Delta \mathbf{x}| K_0(m|\Delta \mathbf{x}|)
	+
	m
	K_1(m|\Delta \mathbf{x}|)
	\right)
\end{align}
Using the fact that
\begin{align}
	\frac{\partial}{\partial x}
	K_0(x) = -K_1(x)
	\qquad
	\frac{\partial}{\partial x}
	K_1(x)
	=
	-\frac{1}{2}
	(K_0(x)+K_2(x))
\end{align}
we end up with
\begin{align}
	2\pi
	\frac{m}{\kappa}
	\lambda_j  
	\frac{\Delta \mathbf{x}^j}{|\Delta \mathbf{x}|} 
	\left(
	\frac{m}{2} 
	K_0(m|\Delta \mathbf{x}|)
	-
	m^2
	|\Delta \mathbf{x}| 
	K_1(m|\Delta \mathbf{x}|)
	-
	\frac{m}{2}
	K_2(m|\Delta \mathbf{x}|))
	\right)
\end{align}
For small values of $|\Delta \mathbf{x}|$ one can show that 
\begin{align}\label{eq120}
	\frac{m}{2} 
	K_0(m|\Delta \mathbf{x}|)
	-
	m^2
	|\Delta \mathbf{x}| 
	K_1(m|\Delta \mathbf{x}|)
	-
	\frac{m}{2}
	K_2(m|\Delta \mathbf{x}|))
	\approx
	-\frac{1}{m |\Delta \mathbf{x}|^2}
\end{align}
while for large values of $|\Delta \mathbf{x}|$ one has
\begin{align}
	\frac{m}{2} 
	K_0(m|\Delta \mathbf{x}|)
	-
	m^2
	|\Delta \mathbf{x}| 
	K_1(m|\Delta \mathbf{x}|)
	-
	\frac{m}{2}
	K_2(m|\Delta \mathbf{x}|))
	\approx
	-m^{\frac{3}{2}}
	\sqrt{\frac{\pi}{2}}
	e^{-m|\Delta \mathbf{x}|}
	\sqrt{|\Delta \mathbf{x}|}
\end{align}
Notice that these contributions to the right hand side of \eqref{nnlstep} are multiplied by the factor $m/\kappa$ and therefore are extremely small. Also, it should be stressed that there are no short distance singularity in \eqref{eq120} since the $\kappa$-deformed theory represents the regime of semiclassical quantum gravity and presumably is to be replaced by complete theory of quantum gravity at the scales smaller than the Planck scale.

We will discuss the properties of the two-point function in more details in a forthcoming paper. For our present purposes it  suffices to conclude that it is the unusual behavior of boost in spacetime representation that is responsible for the violation of the Jost-Wightman-Greenberg theorem in the $\kappa$-deformed context.

\section{Phenomenology of CPT deformation}\label{Sect6}

In this section we incorporate results obtained in Sec. \ref{SecIV} providing us with explicit expressions for the action of the charge conjugation operator $\cal C$ on four-momenta of particles and antiparticles.
We will investigate experimental measurability of deformation at very high energies by observing possible differences in their decay widths, extending results of Ref. \cite{ak-gw_2019} by a more systematic study.

In order to incorporate Eqs. (\ref{eqn:SofP}) to experimental setups let us rewrite them in terms of energies and three-momenta, $E$ and $\mathbf p$, as:
\begin{eqnarray}\label{SofEP}
-S(E) & = &  \sqrt{\kappa^2+M^2}-\frac{\kappa^2}{E+\sqrt{\kappa^2+M^2}} \nonumber \\ & = & E-\frac{\mathbf p^2}{\kappa}
+{\mathcal O}(1/\kappa^2), \ \nonumber \\
-S(\mathbf p)_i & = & \frac{(\mathbf p)_i\kappa}{E+\sqrt{\kappa^2+M^2}} \nonumber \\
& = & (\mathbf p)_i-\frac{(\mathbf p)_i E}{\kappa} + {\mathcal O}(1/\kappa^2),
\end{eqnarray}
where $M$ stands for mass of a particle and its antiparticle, both represented by plane waves.
For an observer boosted with respect to their rest frames with a boost parameter $\gamma$ the antiparticle's energy $\bar S(E)$ differs from that of particle's $E$ by $\mathbf{p}^2/\kappa$.
This difference reflects on different Lorentz factors affecting their lifetimes when passing from the rest frame to the laboratory frame.
The Lorentz boosts are equal to $E/M$ for the particle and $S(E)/M$ for antiparticle.
The time evolution of one-particle unstable states is governed by complex energy eigenvalues $E-\frac i2\Gamma$, where $\Gamma=1/\tau$ stands for decay width being reciprocal of the lifetime $\tau$.
The free, unstable particle and antiparticle plane waves read
\begin{eqnarray}\label{psi1}
\psi_{\mbox{\scriptsize part}}(t) & = & A(M,\Gamma,E) \, \exp \, \left[-i(E-\frac{i}{2}\Gamma) \frac{E}{M} \,t\right], \nonumber \\
\psi_{\mbox{\scriptsize apart}}(t) & = & A[M,\Gamma,S(E)] \, \exp \,\left[-i\left(S(E)-\frac{i}{2}\Gamma\right) \frac{S(E)}{M} \,t\right],
\end{eqnarray}
where $t$ denotes their proper time and $A$ stands for the normalization factor.
Probability density functions for their decays, ${\cal P}=|\psi|^2$, are equal to
\begin{eqnarray}
{\cal P}_{\mbox{\scriptsize part}}(t) & = & \frac{\Gamma E}M\exp \,\left(-\Gamma \,\frac {E}{M}\, t \right),  \label{decay}\\
{\cal P}_{\mbox{\scriptsize apart}}(t) & = & \frac{\Gamma S(E)}{M}\, \exp \, \left[-\Gamma \, \frac{ S(E)}{M}\, t\right] \nonumber \\
                                    & = & \Gamma\left(\frac EM - \frac{\mathbf p^2}{\kappa M}\right) \,\exp\,\left [-\Gamma \, \left(\frac EM - \frac{\mathbf p^2}{\kappa M}\right)\, t\,\right ].\label{decaya}
\end{eqnarray}
It is important to notice that the proper time $t$ in Eqs. (\ref{psi1} - \ref{decaya}) has the interpretation of time that passed after the unstable particle was created.
Therefore, strictly speaking it is a time interval, and expressions are meaningful only for $t\ge 0$.

In the formulas above we assumed that the decay width $\Gamma$ is the same for particles and antiparticles.
This assumption could be challenged on the grounds that we do not know the precise form of the deformed dynamics. However, as we have seen above, the difference between particles and antiparticles in free theory is purely kinematical and we assume that it is possible to construct the interacting theory having the same property.
But then the possible leading-order corrections to the standard decay probabilities will be still of the form (\ref{decay}, \ref{decaya}) and the upper bounds on the value of the deformation parameter $\kappa$ will be valid (up to the order-one parameter describing the relative relevance of kinematical and dynamical effects).

One may also imagine another possibility that the deformation of CPT may get washed up by uncontrollable behaviour of $\Gamma$.
Let us argue that this cannot be the case.
It is indeed very much possible that the decay widths are different for particles $\Gamma$ and antiparticles $\bar\Gamma$, i.e. $\Gamma\neq \bar\Gamma$. The leading-order difference between the two will be of order $1/\kappa$, but since both $\Gamma$ and $\bar\Gamma$ are computed in the rest frame, this difference presumably will be of the order of $m^2/\kappa$ and is negligible.
Since the decay width results from the one-particle irreducible insertions in the propagator, it is conceivable that the difference is of order of $\Lambda^2/\kappa$, where $\Lambda$ is the ultra-violet cutoff, far below the Planck energy, so that also in this case the correction is very small.
Be it as it may, the fact is that the difference between $\Gamma$ and $\bar\Gamma$ {\em does not} depend on momenta of the moving particles and antiparticles, and therefore performing measurements at different energies one can establish the difference between particles and antiparticles even if their decay widths are slightly different. For these reasons we assume that the kinematical effect considered here is dominant at high energies, and in the rest of the paper we keep the decay width of particles and antiparticles equal. 

The qualitative arguments presented above are fully confirmed by a simple calculation presented in  Appendix \ref{Appendix}.

Differences between ${\cal P}_{\mbox{\scriptsize part}}$ and ${\cal P}_{\mbox{\scriptsize apart}}$ in Eqs. (\ref{decay}, \ref{decaya}) can be experimentally tested using precision measurements of lifetimes of particles and antiparticles in experiments at very high energies.
The equality of the total lifetimes of particles and antiparticles is a direct consequence of the CPT theorem (cf. Ref. \cite{Streater:1989vi}).
In case of the partial decay widths, i.e. taking into account uncomplete sets of decay final states, the equality between those of particle and antiparticle is only ensured for final states which can be mixed together in forming eigenstates of the strong part of the decay Hamiltonian.
For the weak and electromagnetic parts of the Hamiltonian such equality is valid only to the lowest order \cite{Sozzi:2008,Lee:1981}.
In this paper we consider the total decay widths and therefore, again, $\Gamma$ in Eqs. (\ref{psi1} - \ref{decaya}) is the same for the particle and antiparticle.
Any difference between particle and antiparticle can originate in a moving frame from deformation of the Lorentz boost $S(E)/M$ for one of them.

Natural candidates for particle and antiparticle lifetimes measurements are the particle-antiparticle pairs, where both objects in a pair are their mutual CPT images, originating from two-body decays of resonances produced on existing or planned accelerators: the Large Hadron Collider (LHC) and the Future Circular Collider (FCC) \cite{fcc}, both at CERN. Such experimental setting is advantageous for our study. Kinematics ensures that in the absence of deformation both particles have the same energies and fly back-to-back with equal momenta in the resonance rest frame. Transforming both particles to the laboratory frame and, ideally, choosing pairs with the same transverse momenta one selects samples suitable for this study. Since an undeformed ${\cal CPT}$ operator leaves energies and three-momenta intact, testing CPT invariance using pairs of CPT-coupled particles requires that their momenta are equal. Particles in pairs originating from resonance decays have non-zero transverse momenta such that even boosted using large Lorentz $\gamma$, their momenta remain slightly divergent with small opening angles $\theta$, of the order $10^{-4}-10^{-6}$ rad for energies considered here.
In order to make the pair strictly CPT-invariant and thus get rid of any differences besides those of the deformation, the momenta should be aligned by rotating them by $\pm\theta$. This is equivalent to assuming neither CPT- nor Lorentz symmetry violation due to anisotropy of space. The latter hypothesis is well supported by experimental results on search for a possible sidereal-time-dependence of interference of neutral $K^0$, $D^0$ or $B^0$ mesons, where no existence of an absolute direction in space has been detected with accuracy comparable or better than considered here, and with much larger rotations in space \cite{Babusci:2013gda,Babusci:2021,lhcb_cpt,babar_cpt,d0_cpt,focus_cpt}.
Event-by-event control of $\theta$ required for direct comparison of momenta with accuracy comparable to ${\mathbf p}^2/(M\kappa)$ is beyond reach of current experimental techniques.
Sufficient statistical accuracy of the mean value of $\theta$ can be only achieved using enormously large data samples.

In Tab. \ref{Table:tab1} we present data on several possible decay channels, Lorentz boosts for energies attainable at LHC and FCC, experimental errors and limits on deformations $\kappa$. Crucial for this study is to have very high energy and not too heavy particles, since $1/\kappa$ corrections to lifetimes are proportional to $\mathbf p^2$ and to the reciprocal of mass. Concerning experimental accuracy, the critical quantity is an overall uncertainty of the particle lifetime $\tau$ that limits sensitivity of the measurement of the actual value of $\kappa$.

In Ref. \cite{ak-gw_2019} we presented results of our study for pairs of muons and possible limitations on $\kappa$ for the LHC and FCC energies. 
It is worthwile to provide a forecast of limits not only for the lightest particle-antiparticle pairs with the most accurately known lifetimes, i.e. $\mu^+\mu^-$, but also for heavier decay products. As seen from Tab. \ref{Table:tab1}, limits on $\kappa$ obtained from the muons and pions are of the same order ($10^{14}$ GeV for LHC and $10^{16}$ GeV for FCC) whereas for tau leptons and mesons containing heavy quarks they are only an order of magnitude weaker. Lighter unstable parent resonances with masses around or below 1 GeV, like $\phi(1020)^0$, $\rho(770)^0$, $\omega(782)^0$ or $K_S$, can be more copiously produced in primary collisions and thus provide better statistical accuracy. Heavier resonances, like $D^0$, $B^0$, $\psi(3770)$ and heavy bottomonia $\Upsilon(10580)$ and $\Upsilon(10860)$, may decay into strange, charm or beauty mesons that live significantly shorter than muons or pions. As seen in Tab. \ref{Table:tab1}, for ultrarelativistic energies the $\gamma$ factor can be as large as $10^3-10^4$ at LHC and $10^4-10^5$ at FCC. This means that for a long-living particle, e.g. muon with $\tau=10^{-6}$ s, an accurate determination of particle's lifetime in laboratory requires an experimental baseline to be very long since $\gamma c\tau = 10^5-10^6$~m at LHC and $10^6-10^7$~m at FCC energies. For short-lived particles, like $B^\pm$ living $10^{-15}$ s, even large Lorentz boosts correspond to $\gamma c \tau$ not exceeding centimetres which could make the measurements easier.

\onecolumngrid

%\pagebreak
%\vspace{.5cm}

\begin{table}[!htbp]
	\caption{\em Limits on deformation parameter $\kappa$ for set of particle-antiparticle pairs and energies attainable at LHC and FCC. Values of decay times with errors and particle masses are from Ref. \cite{pdg}. In calculating these limits for $\kappa$ the lifetime accuracies were assumed everywhere to be $\frac{\sigma_\tau}{\tau}=10^{-6}$. Lorentz boosts $\gamma$ were calculated for energies 6.5 TeV for LHC and 50 TeV for FCC.}
	\label{Table:tab1}
	\vspace{5mm}
 \begin{adjustbox}{angle=90}
 % \begin{tabular}{ c | c | c | c | c | c | c | c | c | c |}
		\begin{tabular}{ p{12mm} p{20mm} p{20mm} p{20mm} p{24mm} p{20mm} p{20mm} p{20mm} p{20mm} p{20mm}}
		\hline
		\addlinespace[2mm]
		\scriptsize Particle & \makecell{\scriptsize Parent \\ \scriptsize resonance} & $\hskip 10pt \tau$[s] & $M$ [GeV] & $\hskip 10pt\frac{\Gamma}{M}$ & \makecell{$\frac{\sigma_\tau}{\tau}$ \\ \scriptsize (from PDG)} & \makecell{$\gamma$ \\ \scriptsize (LHC)}   & \makecell{$\gamma$ \\ \scriptsize (FCC)}  & \makecell{$\kappa=\frac{p^2}{M\delta_\tau}$ \\ \scriptsize(LHC)} & \makecell{$\kappa=\frac{p^2}{M\delta_\tau}$ \\ \scriptsize (FCC)} \\
		\addlinespace[2mm]
		\hline
		
		\addlinespace[2mm]
		$\mu^\pm$ & $J/\psi, \Upsilon$ & $2.2\times 10^{-6}$ & $\hskip 10pt 0.11$ & $2.8\times 10^{-18}$ & $1\times 10^{-6}$ & $6.1\times 10^4$ & $4.7\times 10^5$ & $4\times 10^{14}$ & $2\times 10^{16}$\\
		\addlinespace[2mm]
%		\hline
		
		\addlinespace[2mm]
		$\tau^\pm$ & $J/\psi, \Upsilon$ & $2.9\times 10^{-13}$ & $\hskip 10pt 1.8$ & $1.3\times 10^{-12}$ & $1.7\times 10^{-3}$ & $3.6\times 10^3$ & $2.8\times 10^4$ & $2.5\times 10^{13}$ & $1.5\times 10^{15}$\\
		\addlinespace[2mm]
%		\hline
		
		\addlinespace[2mm]
		$\pi^\pm$ & \makecell{$K_S, \rho^0, \omega^0$ \\ $D^0, B^0$} & $2.6\times 10^{-8}$ & $\hskip 10pt 0.14$ & $1.8\times 10^{-16}$ & $1.9\times 10^{-4}$ & $4.6\times 10^4$ & $3.6\times 10^5$ & $3\times 10^{14}$ & $1.8\times 10^{16}$\\
		\addlinespace[2mm]
%		\hline
		
		\addlinespace[2mm]
		$K^\pm$ & $\phi^0, D^0, B^0$ & $1.2\times 10^{-8}$ & $\hskip 10pt 0.49$ & $1.1\times 10^{-12}$ & $1.6\times 10^{-3}$ & $1.3\times 10^4$ & $1.0\times 10^5$ & $8.5\times 10^{13}$ & $5.1\times 10^{15}$\\
		\addlinespace[2mm]
%		\hline
		
		\addlinespace[2mm]
		$D^\pm$ & $\psi, B^0$ & $1.0\times 10^{-12}$ & $\hskip 10pt 1.9$ & $3.4\times 10^{-13}$ & $6.7\times 10^{-3}$ & $3.5\times 10^3$ & $2.7\times 10^4$ & $2.2\times 10^{13}$ & $1.3\times 10^{15}$\\
		\addlinespace[2mm]
%		\hline
		
		\addlinespace[2mm]
		$B^\pm$ & $\Upsilon$ & $1.6\times 10^{-15}$ & $\hskip 10pt 5.3$ & $0.8\times 10^{-13}$ & $2.4\times 10^{-3}$ & $1.2\times 10^3$ & $0.9\times 10^4$ & $0.8\times 10^{13}$ & $0.5\times 10^{15}$\\
		\addlinespace[2mm]
		\hline
		
	\end{tabular}
\end{adjustbox}
\end{table}

The best experimental accuracy of a mean lifetime, $10^{-6}$, is achieved for muons. The most significant contribution to this number comes from a dedicated measurement at low energy where despite the very large event sample the total error is dominated by the number of events: the statistical error is larger by a factor 3 than the systematic one \cite{tischchenko_2013}. For other particles discussed here: $\tau^\pm, \pi^\pm, K^\pm, D^\pm$ and $B^\pm$, present experimental accuracy is two or three orders of magnitude worse but, if needed, can be improved in currently working flavour factories like BELLE-II, BES-III and experiments at LHC. Therefore, in our estimates we assumed the relative lifetime accuracy of $10^{-6}$ for each particle.

Currently used experimental techniques provide excellent time resolution. For instance, in the spectrometer used by LHCb it amounts to 45 fs for wide spectrum of momenta \cite{lhcb_detector} thus ensuring that systematic contribution to measurements of lifetimes of charged particles should be negligible compared to the statistical one.

A clear distinction between particle and antiparticle is possible due to their decays by using the charge of decay products. In most cases these are leptonic, semileptonic or hadronic decays where charge of the final-state lepton or hadron is equal to that of its parents, e.g. $\mu^\pm\rightarrow e^\pm\nu_e\nu_\mu$, $\pi^\pm\rightarrow\mu^\pm\nu_\mu$, $K^\pm\rightarrow l^\pm\nu_l$ ($l$ standing for $e$ or $\mu$), $K^\pm\rightarrow \pi^\pm X$, etc. If not all decays can be easily identified for experimental reasons, like worse identification or reconstruction efficiency for some particles, one has to take partial widths and modify decay times accordingly.
In Tab.~\ref{Table:tab1} we do not list all potentially interesting decay channels. We omit very rare ones, e.g. $J/\psi(2S)\rightarrow n\bar n, \Lambda^0\bar\Lambda^0, K^+K^-$, or those with neutral final states where the distinction between the particle and antiparticle would render it additionally demanding, e.g. if the final states do not contain CPT-coupled particles, as for $\phi^0\rightarrow K_L K_S$, or require more complex identification of neutral final-state particles via decay chains, as for example for $\psi(3770)\rightarrow D^0\bar D^0$, $\Upsilon\rightarrow B^0\bar B^0$ or $\Upsilon(10860)\rightarrow B_s\bar B_s$.

More discussion is needed in case of the scalar and pseudoscalar neutral mesons. Although the description of $\kappa$-deformed neutral fields is not yet completely developed, being currently under investigations we could present a plausible scenario pertaining to this case. 
On the other hand, in our theoretical investigations in Chapter IV we did not assume that fields are charged and therefore the model is general enough to accomodate neutral fields.
In the conventional quantum field theory, states representing neutral particles are eigenstates of the ${\cal CPT}$ operator. This is no longer going to be the case, however, for the deformed CPT transformation operator which we denote with ${\cal CPT}_\kappa$.  Being mutually their own antiparticles, states like e.g. $|\pi^0\rangle$ and its $\kappa$-deformed image ${\cal CPT}_\kappa \,|\pi^0\rangle$ must refer to different physical objects, which are however indistinguishable in their rest frame. As discussed above, these particles can be distinguished kinematically when transformed to the moving frame, since they have different momenta. In particular, for a $2\,\pi^0$ final state, created e.g. in the decay $K_S\rightarrow \pi^0\pi^0$, this state consists of two $\pi^0$s being their mutual ${\cal CPT}_\kappa$ images with four-momenta $p$ and $ S(p)$, viz. $|\pi^0(p),\pi^0( S(p))\rangle$. Subsequent $\kappa$-deformed transformation ${\cal CPT}_\kappa$ exchanges these two pions and restores their momenta, ${\cal CPT}_\kappa\,|\pi^0(p),\pi^0( S(p))\rangle = |\pi^0( S(p)),\pi^0(p)\rangle$ (up to a phase factor).
This is because ${\cal CPT}_\kappa$ operator is involutive, similarly to the undeformed ${\cal CPT}$ operator, which means that ${\cal CPT}_\kappa({\cal CPT}_\kappa)=\hat 1$ and $ S( S(p))=p$.
Therefore pairs of particles being identical CPT-eigenstates can be also used for finding limits on $\kappa$, similarly to charged ones.
In particular, predictions for $\kappa$ from $K_S\rightarrow \pi^0\pi^0$ decay are close to that for the $K_S\rightarrow\pi^+\pi^-$, although for the neutral decay channel of $K_S$ its lifetime is twice that big and its relative accuracy is twice worse, compared to the charged channel.

Let us now elaborate on another important point of our analysis. As noticed long time ago in Ref. \cite{khalfin}, decay laws have to account for non-zero widths of mass distributions arising from the renormalized propagator of decaying particle. Thus the time- and momentum-dependent decay amplitude $a(t,\mathbf p)$ is given by a distribution $\omega(m; M,\Gamma)$ of the resonance mass $m$, with a mean mass $M$ and decay width $\Gamma$:
\begin{eqnarray}\label{a1}
a(t,\mathbf p)=\int_{-\infty}^\infty dm\,\omega(m; M,\Gamma)\,e^{-it\sqrt{m^2+\mathbf p^2}},
\end{eqnarray}
and the decay probability is equal to $\mathcal{P}(t,\mathbf p)=|a(t,\mathbf p)|^2$.

In the non-relativistic approximation, the resonance mass distribution is approximated by the Breit-Wigner distribution
\begin{eqnarray} \label{a2}
\omega(m; M,\Gamma)= \frac{\Gamma}{2\pi}\frac{1}{(m-M)^2+(\Gamma/2)^2},
\end{eqnarray}
whereas in the more general, relativistic case it is given by
\begin{eqnarray} \label{a3}
\omega(m; M,\Gamma) & = & \frac{f(M,\Gamma)}{(m^2-M^2)^2+M^2\Gamma^2},
\end{eqnarray}
where $f(M,\Gamma)=\frac{2\sqrt{2}}{\pi}\frac{M\Gamma\sqrt{M^2(M^2+\Gamma^2)}}{\big[M^2+\sqrt{M^2(M^2+\Gamma^2)}\big]^{1/2}}$ and does not depend on $m$.
Some refinements of $\omega$ may be considered in the vicinity of thresholds for certain decay channels \cite{fonda_1978} but this could only negligibly affect our main findings here and we do not elaborate on them.
Since we are concerned here with ultrarelativistic particles, our discussion focuses on the relativistic Breit-Wigner distribution (\ref{a3}). Accounting for the non-trivial antipode $S(E)$ and using pole at $m^2=M^2+iM\Gamma$, one gets for $\omega$ given by Eq.(\ref{a3})
\begin{eqnarray} \label{a4}
a(t,\mathbf p) & = & \int dm\,\frac{f(M,\Gamma)}{(m^2-M^2)^2+M^2\Gamma^2} e^{-it(\sqrt{m^2+\mathbf p^2}-\mathbf p^2/\kappa)} \nonumber \\
	& = & e^{-it(\sqrt{M^2+\mathbf p^2+iM\Gamma}-\mathbf p^2/\kappa)}.
\end{eqnarray}
The new decay width $\tilde \Gamma$ can be calculated as imaginary part of the exponent's argument in Eq. (\ref{a4}).
Since the $\kappa$-dependent term contributes only to the real part of it, $\tilde\Gamma$ depends on $M$ and $\Gamma$ and the momentum $\mathbf p$ but not on $\kappa$:
\begin{eqnarray} \label{a5}
\tilde\Gamma & = & 2\,\Im \Big(\sqrt{M^2+\mathbf p^2+iM\Gamma}-\frac{\mathbf p^2}{\kappa}\Big) \nonumber \\
& = & \sqrt{2} \Big[\sqrt{(M^2+\mathbf p^2)^2+M^2\Gamma^2}-(M^2+\Gamma^2) \Big]^{1/2}.
\end{eqnarray}
It has to be clarified here that independence of $\tilde\Gamma$ of $\kappa$ in Eq. (\ref{a5}) only means a lack of specific, deformation-dependent corrections to the decay width from using the Breit-Wigner distributions of the decaying particle mass instead of its sharp value.
An interesting question is if the independence of the decay width of $\kappa$ features more general theory where renormalized propagator is calculated in deformed theory of interactions.
The argument of exponent in amplitude (\ref{a4}) would depend on the location of pole of the renormalized propagator and in particular on its possible dependence on $\kappa$, unknown at present.
On the other hand, the correction term $\mathbf p^2/(2m\kappa)$ to $\Gamma$ in Eq. (\ref{decay}) is determined from the experimentally measured energy and momentum of the particle transformed using deformed ${\cal CPT}_\kappa$ and it is always present in our model.

The formula (\ref{a5}) for the momentum- and mass-dependent decay width can be further simplified by using the smallness of the $\Gamma/M$ ratio, the latter amounting to the order $10^{-12}$ for $\tau^\pm$ and $10^{-16}$ for $\pi^\pm$ and $\mu^\pm$ (cf. Table \ref{Table:tab1}).
Expanding Eq.~(\ref{a5}) to linear terms in $\Gamma/M$ one gets
\begin{eqnarray} \label{a6}
\tilde\Gamma = \frac{M\Gamma}{\sqrt{M^2+\mathbf p^2}} + {\mathcal O}(\Gamma^2/M^2),
\end{eqnarray}
meaning that time in a moving frame is delayed by the Lorentz factor $\gamma=\sqrt{M^2+\mathbf p^2}/(M\Gamma)$.
For the non-relativistic Breit-Wigner distribution, formula analogical to Eq.~(\ref{a5}) and giving the same linear $\Gamma/M$ approximation was obtained in Ref.~\cite{giacosa}.
This approximation is valid with good accuracy since $\Gamma/M$ is of the order $10^{-12}$ in the worst case of $\tau^\pm$ and $K^\pm,$ and $10^{-18}$ in the best case of $\mu^\pm$. It can be checked that terms proportional to $\Gamma^2/M^2$ scale like $1/\gamma^4$ thus giving corrections to $\tilde\Gamma$ not exceeding $10^{-12}-10^{-16}$ for LHC and $10^{-16}-10^{-20}$ for FCC. This means that {\it possible contributions to the decay time coming from non-exponential corrections to the decay law are by far smaller than present experimental accuracy} and thus we do not need to take them into account in our analysis.

\section{Conclusion}

In this paper we considered the $\kappa$-deformed free complex scalar field theory and its symmetry properties. We found that this theory possess Poincar\'e symmetry in the sense that it has ten conserved charges satisfying Poincar\'e algebra. However, as we showed explicitly the boost symmetry acting in space time differs from the undeformed one by the presence of an additional translational factor, different for positive and negative energy modes. We than show that the boost generator does not commute with the operator corresponding to charge conjugation $\cal C$ constructed in this $\kappa$-deformed model. As a consequence the particle and antiparticle with the same rest mass,  are no longer related by charge conjugation when boosted. This leads to the subtle violation of CPT symmetry. We discuss in details the Jost-Wightman-Greenberg theorem that relates Poincar\'e and  CPT symmetry and we show that this theorem fails in the $\kappa$-deformed case as a result of the unusual properties of the boosts.

Violation of CPT symmetry in our model of the $\kappa$-deformed free fields may have observational consequences. In this paper we derived various bounds that can be set on the deformation parameter $\kappa$ describing Planck-scale relativistic kinematics as characterized by the $\kappa$-Poincar\'e algebra using precision measurements for lifetimes of particles {\it vs.} antiparticles. We found the best bound from current data to be $\kappa  \gtrsim (3-4)\times 10^{14}$ GeV obtained for measurements of lifetimes of muons and pions. Using future planned facilities this limit can be pushed up to $\kappa  \gtrsim 2\times 10^{16}$ GeV. We pointed out that there is room for improvement in the sensitivity for measurements of other types of particles. Possible progress in detection techniques leading to better time resolution can move the limits on $\kappa$ even further. Particularly interesting would be imaging technologies based on femtosecond lasers that could hopefully improve time resolutions to be $\sim 1$ fs \cite{chen_2017}.

\section*{Acknowledgment}
The work of WW is supported by the Polish National Science Centre project number 2017/26/M/ST2/00697.  For AB and  JKG, this work was supported by funds provided by the Polish National Science Center,  the project number  2019/33/B/ST2/00050.

\appendix
\section{Estimation of deformed decay width}\label{Appendix}

In this Appendix we present a simple calculation showing that corrections to the decay width resulting from $\kappa$-deformation are negligible. For simplicity we consider only the $\phi\chi^2$ interaction  leaving the more realistic cases to forthcoming publications.

We consider decay of a single $\phi$ particle of mass $M$ to two $\chi$ particles of the mass $m$. We have
\begin{align}
	d\Gamma = \frac{1}{2M} d\,\text{LIPS}_2 |\mathcal{M}|^2
\end{align}
where at tree level $\mathcal{M}=g$ where $g$ is the coupling constant. 
Since $\mathcal{M}$ is computed in the rest frame, the corrections might be only of the form $m/\kappa$ and $M/\kappa$. In the {\em undeformed theory} the Lorentz-invariant phase space factor has the form
\begin{align}
	d\,\text{LIPS}_2
	&=
	(2\pi)^4 \delta^4(k_1+k_2 - k)
	\frac{d^3 k_1}{(2\pi)^3 2\omega_{k_1}}
	\frac{d^3 k_2}{(2\pi)^3 2\omega_{k_2}} \nonumber\\
	&=
	\frac{1}{4(2\pi)^2\omega_{k_1}\omega_{k_2}} \delta^4(k_1+k_2 - k)
	d^3 k_1
	d^3 k_2\label{dGamma},
\end{align}
where $k$ is the incoming momentum, while $k_1, k_2$ are the outgoing ones.

In the deformed case the only difference comes from the fact that instead of the ordinary delta function we have the deformed momentum composition rule
\begin{align}
	\delta^4(k_1 \oplus k_2 \oplus S(k))
\end{align}
Since here we want to find just the leading order deviation from the undeformed theory, we consider only one ordering here; however one can show that all other orderings will give the same first-order result. 

Before continuing let us recall that to the leading in $1/\kappa$ it follows from \eqref{II.5}, \eqref{II.6} that
\begin{align}
(p\oplus q)_0 &= p_0+q_0 +\frac{\mathbf{p}\mathbf{q}}\kappa\,,\quad (p\oplus q)_i= p_i+q_i +\frac{p_i\, q_0}\kappa \nonumber\\
S(p)_0 &= - p_0 + \frac{\mathbf{p}^2}\kappa\,,\quad S(p)_i = - p_i +\frac{p_ip_0}\kappa
\end{align}

We compute $\Gamma$ in the reference frame in which the initial particle is at rest so that $\mathbf{k}=0$. Then
\begin{align}
	\delta^4(k_1\oplus k_2 \oplus S(k)) &= \delta\left(\omega_{\mathbf{k}_1}\oplus \omega_{\mathbf{k}_2}\oplus S(\omega_{\mathbf{k}})\right)\delta^3\left(\mathbf{k}_1\oplus \mathbf{k}_2\right)\nonumber\\
	&= \delta\left(\omega_{\mathbf{k}_1}+ \omega_{\mathbf{k}_2}- \omega_{\mathbf{k}} + \frac{\mathbf{k}_1\mathbf{k}_2}\kappa\right)\delta^3\left(\mathbf{k}_1+ \mathbf{k}_2+\frac{\mathbf{k}_1\omega_{\mathbf{k}_2}}\kappa \right)
\end{align}
Substituting this to \eqref{dGamma} and integrating we get
\begin{align}
\Gamma
	&=\frac{g^2}{2M}\int
	\frac{d^3 k_1
	d^3 k_2}{4(2\pi)^2\omega_{\mathbf{k_1}}\omega_{\mathbf{k_2}}} 
	\, \delta\left(\omega_{\mathbf{k}_1}+ \omega_{\mathbf{k}_2}- \omega_{\mathbf{k}} + \frac{\mathbf{k}_1\mathbf{k}_2}\kappa\right)\delta^3\left(\mathbf{k}_1+ \mathbf{k}_2+\frac{\mathbf{k}_1\omega_{\mathbf{k}_2}}\kappa \right)\nonumber\\
	&=\frac{g^2}{2M}\int
	\frac{\mathbf{k}_2^2 d k_2
	}{4\pi\omega^2_{\mathbf{k_2}}} 
	\,\left(1+\frac{\mathbf{k}^2_2}{\omega_{\mathbf{k}_2}\kappa}-3\frac{\omega_{\mathbf{k}_2}}\kappa\right)\delta\left(2 \omega_{\mathbf{k}_2}- M - \frac{2\mathbf{k}^2_2}\kappa\right)
	\label{Gammaint}.
\end{align}
The argument of the delta function is zero for
\begin{equation}
  \mathbf  k^2_2 = \left(\frac{M^2} 4 - m^2\right)\left(1-\frac{M}\kappa\right)
\end{equation}
and using this fact to explicitly compute the above integral, one reaches the conclusion that
\begin{align}
	\Gamma
	=
	\Gamma^U
	\left[
	1 + \frac{M}{\kappa}
	\left(
	2\frac{m^2}{M^2}
	+
	\frac{24 m^4}{M^4}
	-
	1
	\right)
	\right]
\end{align}
where $\Gamma^U$ denotes the undeformed $\Gamma$. We see that the kinematical corrections to the integral \eqref{Gammaint} resulting from the deformation are of the form $M/\kappa$ . 
Together with possible corrections to the coupling constant $g$, which are of the same order, we conclude that overall corrections to the decay width $\Gamma$ are at most $m/\kappa$ or $M/\kappa$, i.e. of order $10^{-19}$, and therefore completely negligible. 
Moreover, since even in deformed case the masses of particles and antiparticles are identical, the corrections to the decay width $\Gamma$ are the same for particles and antiparticles, if we start with real interaction Lagrangian.

\end{document}